\documentclass[prd,reprint,amsmath,nofootinbib]{revtex4-1}
\pdfoutput=1

\usepackage{ifthen}
\usepackage{epsfig}
\usepackage{booktabs} 
\usepackage{natbib}
\usepackage{wasysym}
\usepackage{feynmp}
\usepackage{slashed}

\newcommand{\D}{\text{d}}

\newcommand{\beqra}{\begin{eqnarray}}
\newcommand{\eeqra}{\end{eqnarray}}
\newcommand{\beq}{\begin{equation}}
\newcommand{\eeq}{\end{equation}}

\begin{document}

\title{Model independent limits on an ultra-light gravitino from Supernovae}

\author{Riccardo Catena}
\email{riccardo.catena@theorie.physik.uni-goettingen.de}
\author{Laura Covi}
\email{laura.covi@theorie.physik.uni-goettingen.de}
\affiliation{Institut f\"ur Theoretische Physik, Friedrich-Hund-Platz 1, D-37077 G\"ottingen, Germany}
\author{Timon Emken}
\email{timon.emken@physik.uni-wuerzburg.de}
\affiliation{Institut f\"ur Theoretische Physik und Astrophysik, Universit\"at W\"urzburg, Am Hubland, D-97074 W\"urzburg, Germany}

\begin{abstract}
We revisit astrophysical constraints on models with an ultra-light gravitino, in particular extending 
the analysis to more general models and the case of R-parity breaking. These constraints allow to restrict the
value of the gravitino mass depending on the masses of scalar moduli. We perform both a frequentist and
bayesian analysis and we find comparable results, even if the bayesian analysis is in part affected by
volume effects. Only a small window of gravitino masses can be excluded by the SN constraints in
a model-independent way, while limits obtained with a definite assumption on the scalar masses 
result more stringent.
\end{abstract}

\keywords{Supergravity and Supersymmetry, Gravitino Phenomenology, Supernovae} 

\maketitle

\section{Introduction}

Cosmology and astrophysics are among the most successful probes for very weakly interacting particles
within extensions of the Standard Model of Particle Physics.
Indeed a substantial part of our knowledge also about the more weakly interacting Standard Model particles, 
the neutrinos, comes from cosmological and astrophysical data. 

Particularly interesting is the case of the gravitino, the superpartner of the graviton in supergravity
extensions of the Standard Model. For such particle the couplings are all set by the symmetry and
cannot be modified easily. It was therefore realized very early on, that gravitinos can be very dangerous
long-lived relics in the early universe if they are not stable. This is often referred to as the 
{\it gravitino problem} in cosmology \cite{Khlopov:1984pf,Ellis:157091}. Of course the other possibility is for the gravitinos to be
the lightest supersymmetric particle and in that case they can also successfully play the role of
Dark Matter \cite{pagels1982,Bolz:1998ek,Fujii:2003nr,Gorbunov:2008ui,Catena:2013pka}. 
We are here instead interested in a different scenario, where the gravitinos are ultra-light,
with masses much smaller than the electroweak scale or even the other SM particle masses.
In that case gravitinos can be in thermal equilibrium and survive as thermal relics, but they play practically 
no role in cosmology, since their energy density is negligible.
An ultralight gravitino can be naturally obtained in supersymmetric models of the no-scale type and 
gives rise to a very interesting phenomenology thanks to its enhanced couplings \citep{fayet19772,Brignole:1996fn,Brignole:1998uu,Brignole:1997sk,Brignole:1998me,Brignole:1999gf}, as
we will review in sec. \ref{ss:gravitinolagrangian}. In that case, thanks to the large interaction with
normal matter, astrophysics provides the possibility to constrain the scenario. Indeed
astrophysical object like stars and Supernovae are highly sensitive to any additional
cooling mechanism, apart for neutrino emission, and have been exploited in the
past to put limits on new particles, e.g. in the case of axions \cite{Raffelt:2006cw}. 

In the past various approaches to restrict an ultra-light gravitino through astrophysical observations have been made. 
Some early constraints on its mass were derived from cosmology, more precisely from Big Bang Nucleosynthesis (BBN). 
It was shown that BBN allows either a light gravitino $<1$keV or a very heavy one \citep{pagels1982,weinberg1982}. 
These investigations have been revisited in 1993 by Moroi~\citep{moroi1993} and four years later again by 
Gherghetta \citep{gherghetta1997,gherghetta19972}. Other studies investigated the possibility of exotic cooling of stars, 
red giants and white dwarfs \citep{nowakowski1995,fukugita1982}. These limits were reviewed in \citep{grifols1998}, where
it is shown that the limits depend as well on the masses of the scalar partners of the gravitino.

Our analysis is based on \citep{emken2013}, wherein we revisited and generalized some of the results obtained by Grifols, Mohapatra and 
Riotto during the '90s \citep{grifols1996,grifols1997}. They were able to exclude a gravitino mass range covering a few orders of 
magnitude in the ultra-light mass regime from the SN1987A observation. The exact mass interval however depends on the masses of 
the sgoldstinos, the scalar superpartner of the goldstino field, which are model-dependent.  In the early works \citep{grifols1996} such 
fields were implicitly assumed to be massless. 
Moreover most of these results were obtained assuming a specific relation between the sgoldstino's coupling parameters $c$ and $d$ and
the photino and gravitino mass, namely $c\cdot d = \frac{m_{\tilde{\gamma}}}{m_{3/2}}$, where $m_{3/2},m_{\tilde{\gamma}}$ are the 
masses of the gravitino and photino  respectively. However this kind of relation does not have to occur in general models.

We generalize here previous results by performing a full parameter scan in the model-dependent parameters such as the sgoldstino masses 
and couplings. Moreover we consider also possible effects of R-parity breaking, that opens up the possibility of a single-gravitino
production channel.

\section{The Energy Loss Argument and Constraints from Supernovae}
\label{sec:data}
A core collapse of stars with masses larger than a few solar masses can trigger a giant stellar explosion called a Supernova. Its dynamics can be explained by a ``bounce-and-shock'' model\footnote{For more details on SNe and SN cooling we refer to the textbook \cite{raffelt1996}.}. At the end of the star evolution the iron core can no longer release energy via fusion and remains stable as long as the electron degeneracy pressure balances the gravitational pressure. However, once the core mass exceeds the Chandrasekhar limit of $M\approx 1.44 M_{\astrosun}$ it becomes unstable and starts to collapse.  Shortly afterwards, the gravitational binding energy 
\begin{align}
E_b \sim \frac{3}{5}\frac{G_N M}{R_{SN}} = 1.6\times 10^{53}\left(\frac{M}{M_{\astrosun}}\right)\left(\frac{R_{SN}}{10\text{km}}\right)^{-1}\text{erg} \label{eq:bindingenergy}
\end{align}
gets released. A large fraction of this energy is emitted in the form of neutrinos. In Eq.~\eqref{eq:bindingenergy} $R_{SN}\sim 10$~km is the core radius.

On February 23th 1987 the blue giant Sanduleak-69202 in the Large Magellanic Cloud exploded in a Supernova that was named SN1987A. It is the first, and so far the only Supernova, for which the emitted neutrinos have been observed directly. They have been detected by Kamiokande II and IMB \cite{bionta1987,hirata1987}. Models of gravitational collapse \cite{burrows1986,mayle1986} allow to estimate the neutrino energy released by SN1987A as
\begin{align}
E_{\nu}> 2\times 10^{53}\text{erg}\, . \label{eq:neutrinoenergy}
\end{align}
Combining Eqs.~\eqref{eq:bindingenergy} and \eqref{eq:neutrinoenergy}, one finds the upper limit 
\begin{align}
L_X < 10^{52}\frac{\text{erg}}{\text{s}}\,  \label{eq:snconstraint}
\end{align} 
on the luminosity $L_X$ which can be produced by anomalous cooling mechanisms present in SN1987A. Many models for physics beyond the Standard Model predict new light weakly interacting particles which can in principle contribute to $L_X$. 

At this point we mention that constraints from anomalous SN cooling can also been found for other proposed light particles such as axions \citep{raffelt1996,raffelt2006} and sterile neutrinos \citep{kolb1996}. Furthermore it is also possible to apply the energy loss argument to alternative astrophysical settings such as massive stars with $8-10$ solar masses, which has been done for axions in 2012 \cite{friedland2012}. In this paper we focus on Supernovae, since they reach a larger core temperature compared to stars and
are therefore expected to give the strongest constraints. We compute in the following the contribution to $L_X$ generated by ultralight gravitinos production 
via photon collisions during a Supernova of type II and we apply the constrain given in Eq.~\eqref{eq:snconstraint}. 
As derived in Appendix A, the constraint is given as a function of the gravitino production cross-section as
\begin{align}
L_X &>\frac{4V}{(2\pi)^6}\int \D^3 p_1 \D^3p_2\; e^{-(p_1^0+p_2^0)/T}(p_1^0+p_2^0) \frac{p_1 \cdot p_2}{p_1^0 p_2^0} \nonumber\\
&\times \sigma(\gamma\gamma\longrightarrow \widetilde{G}\widetilde{G})\, , \label{eq:ggluminosity1}
\end{align}
where $p_1, p_2$ are the momenta of the incoming photons, $n_\gamma (p_i)$ has been approximated by $ e^{-p_i^0/T} $ and the factor $4$
comes from the two possible photon polarizations.\footnote{This expression differs exactly by the factor $4$ compared to the one used previously in \cite{grifols1996}.}

The described argument applies only if the gravitinos produced at SN1987A escape the Supernova core and carry away energy. If the gravitino-matter coupling is sufficiently strong, the gravitino mean-free-path $\lambda_{\rm mfp}$ could be short enough for the gravitinos to diffuse inside the Supernova core. If the gravitinos are trapped inside the core radius $R_{SN}\sim10$~km for longer than $\sim 1$~s, their energy is depleted by neutrino emission and the gravitino luminosity is again compatible with $L_X<10^{52}\text{erg}/\text{s}$. Trapped gravitinos random-walk through the core and cover a distance $\sim\sqrt{N}\lambda_{\rm mfp}$ in a time interval $\lambda_{\rm mfp} N/c$, where $N$ is the number of gravitino scatterings. Hence, we obtain the lower bound for $\lambda_{\rm mfp}$,
\begin{align}
\lambda_{\rm mfp} &\geq \frac{R_{SN}^2}{c (1\,\text{s})}\approx 0.3\; \text{m}\, .\label{eq:mfpconstraint}
\end{align}
The limit \eqref{eq:snconstraint} applies therefore for gravitino mean-free-paths larger than \eqref{eq:mfpconstraint} only. In this paper we investigate how Eqs.~\eqref{eq:snconstraint} and \eqref{eq:mfpconstraint} constrain the parameter space of the effective theory of ultra-light gravitinos that we introduce in the next section.

\section{Theoretical framework}
\label{sec:theory}
In this section we briefly introduce the theoretical framework assumed in Sec.~\ref{sec:results} deriving limits on the gravitino mass from Supernovae observations. 
Our general results for the cross-sections used in the analysis are given in Eqs.~\eqref{eq:productioncs} and \eqref{eq:scatteringcs}, obtained here for the first time.

\subsection{Gravitino Lagrangian}
\label{ss:gravitinolagrangian}

We start here with the effective Lagrangian for a locally supersymmetric version of QED with broken SUSY and conserved R-parity, which was first
derived in \cite{cremmer1983,bhattacharya1988}.  It inherits a canonical K\"ahler potential, a vanishing cosmological constant and has been studied 
by various authors  in the context of gravitino phenomenology \cite{bhattacharya1987,bhattacharya1988,gherghetta1997,gherghetta19972,grifols1996,grifols1997}. 
The matter fields $\chi_L^i$ and $\phi^i$, the gauge fields $A_{\mu}$ (of field strength $F_{\mu\nu}$) and $\lambda$, and the gravitino $\psi_\mu$, as well as their 
interactions are described by the following Supergravity Lagrangian~\footnote{
Throughout this paper we employ the conventions established in \cite{peskin1995}, notably we choose $(+---)$ as metric signature, 
fix the sign of the Levi-Civita symbol via $\epsilon_{0123} =-1$ and set $c=\hbar=k_B=1$.},
\begin{align}
e^{-1}\mathcal{L}&= -\frac{M_P^2}{2} R  -  \frac{1}{2}e^{-1} \epsilon^{\kappa\lambda\mu\nu}\overline{\psi}_{\kappa}\gamma^5\gamma_{\lambda} D_{\mu}\psi_{\nu} \nonumber\\
&+  \frac{i}{2} \, m_{3/2}  \overline{\psi}_{ \alpha }^{}\sigma^{\alpha\beta}\psi_{\beta}-\frac{1}{4}  F_{\mu\nu}F^{\mu\nu}\nonumber\\
&+ \frac{i}{2} \overline{\lambda}^{(a)}\left[ \gamma^{\mu}D_{\mu} - m_{\tilde{\gamma}}\right]\lambda^{(a)}+  D_{\mu}\phi^{i}D^{\mu}\phi^{*i} - m_{\phi_i}\phi^{*i}\phi^i \nonumber\\
&+ i \overline{\chi^i}_L \gamma^{\mu}D_{\mu}\chi_L^i - \frac{1}{2}m_{\chi_i} \left( \overline{\chi^c_L}^i \chi_L^i + \text{h.c.}\right) \nonumber\\
&- \frac{i}{\sqrt{2} M_P}\left( D_{\mu}\phi^{*i}\overline{\psi}_{\nu}\gamma^{\mu}\gamma^{\nu}\chi^i_L - D_{\mu}\phi^i \overline{\chi}^i_L \gamma^{\nu}\gamma^{\mu}\psi_{\nu}\right) \nonumber\\
&- \frac{1}{4 M_P}\overline{\psi}_{\mu}\sigma^{\rho\sigma}\gamma^{\mu}\lambda \; F_{\rho\sigma}+\mathcal{O}(M_P^{-2})\, ,\label{eq:modelL}
\end{align}
where  $M_P= 2.2 \times 10^{18} $~GeV is the reduced Planck mass and $R$ is the Ricci scalar. The covariant derivatives are given by 
\begin{align}
D_{\mu}\phi^i &= \partial_{\mu} \phi^i + i Q_i  A_{\mu}\phi^i\,, \nonumber\\
D_{\mu} \chi_L^i &= \partial_{\mu}\chi^i_L + \frac{i}{4}\omega_{\mu ab}\sigma^{ab}\chi_L^i + i Q_i A_{\mu}\chi_L^i\, ,\nonumber\\
D_{\mu}\lambda &= \partial_{\mu}\lambda + \frac{i}{4}\omega_{\mu ab}\sigma^{ab}\lambda\,, \nonumber\\
D_{\mu}\psi_{\nu} &= \partial_{\mu}\psi_{\nu}+\frac{i}{4}\omega_{\mu ab}\sigma^{ab}\psi_{\nu}\, .
\end{align}
Here $Q_i$ is the charge of the field on which the covariant derivative acts, and $\omega^{\mu}_{ab}$ is the spin connection.

The value of the gravitino mass $m_{3/2}$ heavily depends on the SUSY breaking scheme. Certain models, such as no-scale models \citep{ellis1984,cremmer19832,lahanas1986} and models with gauge-mediated SUSY breaking (GMSB) \citep{dine1996,dine1995,nelson1995} allow the gravitino to be very light. As we will see below, a very small value of $m_{3/2}$ is a phenomenologically attractive possibility, since it enhances the gravitino interactions, which are otherwise suppressed by the small gravitational coupling constant $1/M_P$~\citep{fayet19772}.

A massive gravitino obtains its $\pm\frac{1}{2}$ helicity states by absorbing the goldstino field via the Super-Higgs mechanism~\citep{cremmer1983,volkov1973,cremmer1978,nilles1983,deser1977}. If the gravitino mass is very small compared to the energy scale of the relevant 
processes, its $\pm\frac{1}{2}$ helicity states dominate in the transition amplitudes (with some exception discussed below), and the gravitino effectively 
behaves like a massless goldstino. In this case the gravitino $\pm\frac{3}{2}$ helicity states are negligible in the calculations. This result is a consequence of  
the SUSY equivalence theorem \cite{casalbuoni1988}. In the limit in which the equivalence theorem applies, we can approximate the gravitino field $\psi_{\mu}$ as
\begin{align}
\psi_{\mu}\sim i \sqrt{\frac{2}{3}} \frac{1}{m_{3/2}}\partial_{\mu}\eta \, , \label{eq:etrough}
\end{align}
where $\eta $ is the spin-$\frac{1}{2}$ goldstino. The small $m_{3/2}$ term in the denominator of \eqref{eq:etrough} leads to an enhancement of the gravitino interactions. 
In certain models this enhancement of the gravitino interactions affects also the sgoldstinos, the scalar superpartners of the goldstino. In contrast to the goldstino, 
the sgoldstinos do not disappear from the physical spectrum. They can be very light, with masses of the order of the gravitino mass~\citep{ellis19842,dicus1990}, 
or also much heavier. Indeed, light sgoldstino are generally very long-lived and can be problematic for cosmology, so that heavy sgoldstinos are favoured
from that point of view.
In general though, the scenario with an ultralight gravitino could not only allow observations of gravitational effects due to the gravitino, but also due to 
new particles from the hidden SUSY breaking sector.

In using Eq.~\eqref{eq:etrough}, we implicitly neglect terms suppressed by powers of  $m_{3/2}$ higher than $-2$ in the gravitino polarization tensor (see Eq.~\eqref{eq:poltensor2} below). In the case of processes with more than one external gravitino, and involving several diagrams exhibiting different dependencies on $m_{3/2}$, these 
higher order terms are not negligible. Rather than using the complete spin-$3/2$ polarization tensor $\Pi^{\pm}_{\mu\nu}(k)$, given e.g. in \citep{moroi1995}, it is convenient to expand  $\Pi^{\pm}_{\mu\nu}(k)$  in powers of $m_{3/2}$,
\begin{align}
 \Pi^{\pm}_{\mu\nu}(k)  &\equiv  \frac{1}{m_{3/2}^{2}}\Pi^{(2)}_{\mu\nu}(k)\pm\frac{1}{m_{3/2}} \Pi^{(1)}_{\mu\nu}(k)+ \Pi^{(0)}_{\mu\nu}(k) \nonumber \\ & +\mathcal{O}(m_{3/2}) \,,\label{eq:poltensor2}
\end{align}
performing then a careful $m_{3/2}$ power counting in the relevant amplitudes, and identifying so the leading contributions to the observables. In Eq.~\eqref{eq:poltensor2}, 
we  denote by $\Pi^{(n)}_{\mu\nu}(k)$ the coefficient of the $m_{3/2}^{-n}$ term in this power series expansion. With this notation, $m_{3/2}^{-2} \Pi^{(2)}_{\mu\nu}(k)$
corresponds to the polarization tensor resulting from Eq.~\eqref{eq:etrough}.

\subsection{Supersymmetry breaking and the goldstino multiplet}

As we have seen, the scalar partners of the goldstino remain as physical fields in the spectrum and they can be as light
as the gravitino. We can write the general effective sgoldstino Lagrangian as follows
\begin{align}
e^{-1}\mathcal{L}_{\text{sgoldstino}} &=\frac{1}{2}\left( \partial_{\mu} S\partial^{\mu}S -  m_S S^2+\partial_{\mu} P\partial^{\mu}P- m_P P^2\right)\nonumber\\
& +\frac{c}{4 M_P} F_{\mu\nu}F^{\mu\nu}S+i\frac{d}{2M_P}\; m_{3/2} \overline{\psi}_{\mu}\sigma^{\mu\nu}\psi_{\nu}S\nonumber\\
& - \frac{c}{8 M_P} e^{-1} \epsilon^{\mu\nu\rho\sigma}F_{\mu\nu}F_{\rho\sigma}P \nonumber\\ 
& - i\frac{d }{4 M_P} \epsilon^{\mu\nu\rho\sigma}\overline{\psi}_{\mu}\gamma_{\nu}\psi_{\rho}\partial_{\sigma}P+\mathcal{O}(M_P^{-2})\, .\label{eq:lsgoldstino}
\end{align} 
In Eq.~\eqref{eq:lsgoldstino}, the scalar and pseudo-scalar fields, respectively $S$ and $P$, are the real sgoldstino components~\cite{bhattacharya1988} given by
\begin{align}
S=\frac{1}{\sqrt{2}}\left(\phi_S+\phi_S^*\right)\, , \quad P=\frac{1}{\sqrt{2}i}\left(\phi_S-\phi_S^*\right)\, ,
\end{align}
where $\phi_S$ is a scalar component of the $\Phi_S$ chiral multiplet in the hidden sector, whose non-vanishing $F$-term breaks supersymmetry. 
The sgoldstino couplings constants $c$ and $d$ are model dependent, but can be related to other SUSY breaking parameters in specific scenarios 
as we will see below. They  largely depend on the $\Phi_S$-dependent contribution to the gauge kinetic function, on the K\"ahler potential and on the 
superpotential. Also model-dependent are the sgoldstino masses. In fact, even if the sgoldstino direction is univocally determined by the SUSY breaking 
direction in the scalar field space, the scalar mass matrix is often non-diagonal and the sgoldstinos in general mix with the other moduli fields,
so that they are not in general eigenstates of the scalar mass matrix (see e.g. \cite{GomezReino:2006dk, GomezReino:2006wv, Covi:2008ea} for
constraints arising from the scalar mass matrix in Minkowski and de-Sitter vacua).
We will consider the Lagrangian given above to include in an effective way the presence also of those additional scalars. 
Indeed the presence of many scalar can be approximated by an appropriate increase of the 
gravitino-scalar couplings.
In deriving limits on the gravitino mass, we assume the Lagrangian (\ref{eq:lsgoldstino}), with 
$m_S$ and $m_P$, and the product of coupling constants $c\cdot d$, as free parameters.

We now provide examples for how (\ref{eq:lsgoldstino}), and in particular $m_{3/2}$, can be generated in specific 
models for supersymmetry breaking. Though to specify the mechanism of supersymmetry breaking and mediation is not 
necessary in our model-independent study, it can allow to explicitly relate $m_{3/2}$ to the scale of supersymmetry 
breaking and then to the rest of the supersymmetric spectrum.

The mechanism responsible for supersymmetry breaking is yet unknown, but we can consider two general classes of models,
depending if the breaking is mainly due to the superpotential, $F$-term breaking, or the gauge sector, i.e. 
$D$-term breaking.
We will here shortly review the prediction for the mass spectrum and sgoldstino couplings in the first case.
In the case of $D$-term breaking of supersymmetry, in order to obtain a vanishing cosmological constant,
also a non-vanishing superpotential $W$ and therefore $F$-terms are present at the minimum, leading to a 
combined $D$-term and $F$-term breaking, where often the $F$-term dominates~\cite{ArkaniHamed:1998nu, Kawamura:1998gy}. 
If supersymmetry is broken by non-vanishing $F$-terms for some of the chiral multiplets, the goldstino field 
is the corresponding combination of the fermionic partners, i.e. 
$ \eta = \sum_i F_i \chi_L^i/\sqrt{\sum_i |F_i |^2 }$. 
Very often one approximates this picture by assuming that one single 
chiral multiplet breaks supersymmetry and then the goldstino is simply $ \eta = \chi_S $. 

A simple computable model of this type is the Polonyi 
model \cite{Polonyi}, based on a  chiral multiplet $\Phi_S$ in the hidden sector with canonical K\"ahler potential and the superpotential 
\begin{equation}
W_P = M_S^2 (\Phi_S + \beta)
\label{eq:Polonyi}
\end{equation}
where $ M_S $ is the scale of SUSY-breaking, since the $\Phi_S $ F-term is $F_S = M_S^2 $, and the constant $\beta = - (2-\sqrt{3}) M_P $ 
is chosen to ensure zero vacuum energy at the potential minimum. In this simple case the sgoldstinos are just the scalar 
and pseudoscalar components of the $\Phi_S$ superfield as given above and their masses are of the order of the gravitino mass:
\begin{equation}
m_S, m_P \sim m_{3/2} \sim \frac{M_S^2}{M_P}\; .
\end{equation}
In this scenario all the moduli fields of the theory gain similar masses via gravity mediation, giving rise 
to a multidimensional non-diagonal scalar field mass matrix. The goldstino direction is in general not an 
eigenstate of such a matrix and therefore non-trivial mixings can arise in the scalar sector. 
Also the gauginos obtain a mass via gravity mediation of a similar order, if the gauge kinetic function has 
a dependence on $\Phi_S$. 
For an ultralight gravitino such a light mass spectrum is long excluded by collider bounds on the mass 
of the SM superpartners.

Another popular mechanism of mediation, providing a large mass splitting between the SM superpartners and 
the gravitino, is gauge mediation \cite{Giudice:1998bp}. 
This mechanism can be easily embedded in the model~(\ref{eq:Polonyi}) just by adding a coupling of the Polonyi 
field to 
SM charged messengers as $ (\lambda \Phi_S + M_\Phi) \Phi_j \bar \Phi_j $ with $j=1, ... N$~\cite{Martin:1996zb, Hisano:2008sy}. 
Then the gaugino and scalar masses are generated at one and two-loop level respectively and read
\begin{equation}
M_i \sim N\; \frac{\alpha_i}{4\pi} \frac{\lambda M_S^2}{M_\Phi} \quad\quad
m_0^2 = \sum_i C_{R,i} N^2 \left(\frac{\alpha_i}{4\pi} \frac{\lambda M_S^2}{M_\Phi} \right)^2 
\label{eq:m0m12}
\end{equation}
where $C_{R,i} $ are SM representation dependent constants~\cite{Martin:1996zb}. 
We see then that the ratio between the gravitino and the photino mass is given by
\begin{equation}
\frac{m_{\tilde\gamma}}{m_{3/2}} \sim \frac{M_1}{m_{3/2}} \sim N\; \frac{\alpha_i}{4\pi} \frac{\lambda M_P}{M_\Phi} \,.
\label{eq:mphotino}
\end{equation}
Note that in the simplest model, barring cancellations, the mass of the messengers is also given by the vacuum
expectation value of $S$ by the term $ \lambda \langle S \rangle  \sim \lambda M_P $, giving again masses of
similar order. 
Assuming though the presence of more fields $\Phi_S $, it is possible to disentangle the vacuum expectation
value of the scalar field and the F-term and obtain also very different ratios of masses. We see that to have a large 
mass splitting allowing for an ultralight gravitino we are therefore obliged to consider more complicated
hidden sectors and many very light messengers, just around the electroweak scale.

Regarding the couplings between the sgoldstinos and the QED gauge multiplet, they arise in gauge mediation
via a similar one-loop diagram to that generating the gaugino masses, just substituting the F-term insertion with a 
physical scalar field. We therefore expect in the simplest case that the coupling $c$ is given by
\begin{equation}
\frac{c}{M_P} \sim \frac{m_{\tilde \gamma}}{F_S} \rightarrow \quad c \sim \frac{m_{\tilde \gamma}}{m_{3/2}}\;
\end{equation}
but in the more general case cancellations between diagrams or the presence of scalar mixing parameters 
can give rise to smaller values of the couplings. Note that in  gravity mediation the same coupling is instead
$c \sim \sqrt{3/2} \frac{m_{\tilde \gamma}}{m_{3/2}} $ \cite{cremmer1983,bhattacharya1988}.
The couplings $d$ between the gravitino and the sgoldstinos are a remnant of the usual supergravity 
coupling between the gravitino, a scalar field and its superpartners for the goldstino multiplet. In the
limit of light gravitino and single goldstino field, the dominant coupling is then of the form given above with
\begin{equation}
d \sim  \sqrt{\frac{2}{3}}
\end{equation}
while in the case of many scalar moduli  the coupling can be reduced by mixing matrices.
So we obtain for the relevant product of coupling, which will later appear in the cross-sections,
\begin{equation}
\xi \equiv c\; d  \leq \frac{m_{\tilde \gamma}}{m_{3/2}}\; .
\end{equation}

We will take this value as the maximal possible value in the following and consider later also the 
simplified expression obtained for $ m_{\tilde\gamma} \sim 100 $ GeV, i.e.
\begin{equation}
\xi  = \frac{100\;\mbox{GeV}}{m_{3/2}}\; 
\end{equation}
in order to compare with previous works~\cite{cremmer1983,bhattacharya1988,grifols1996,grifols1997}. 
Note that in this case the product of couplings is not an independent parameter, but it is determined 
by the gravitino mass. In the general case, instead, since the scale of $ m_{\tilde\gamma} $ is arbitrary,
$\xi $ is an independent parameter.

\subsection{Supersymmetry breaking models with natural ultra-light gravitino}

In the model discussed above, a single scale, namely $F_S$, determines the gravitino mass via 
$m_{3/2}\sim F_S/M_P$, and the masses of the remaining superpartners through Eqs.~(\ref{eq:m0m12}). 
In the light of existing limits on gaugino and sfermion masses, it can hence be difficult to accommodate 
an ultra-light gravitino within this model, even if it is possible when a very large number of messenger
fields $N$ is considered.

Models where gaugino and sfermion masses are to some extent independent of the gravitino mass are 
of great interest for our analysis, since in that case ultra-light gravitinos can exist avoiding constraints 
from LHC and direct searches for new physics. Within the many possibilities, we mention here a few
examples of this type.

A family of models where ultra-light gravitinos, and TeV gauginos and sfermions can coexist is discussed 
in~\cite{Ellis,Nanopoulos}. The theory assumes a no-scale form for the K\"ahler function, and a diagonal 
non-canonical gauge kinetic function $f\sim \exp(-A S^q)$, where $A$ and $q$ are constants, and $S$ is the 
hidden sector field responsible for supersymmetry breaking. In this case the gaugino and gravitino masses 
are related as follows 
\begin{equation}
M_{i} \sim \left(\frac{m_{3/2}}{M_P}\right)^{1-\frac{2}{3}q} M_P\, ,
\end{equation}
and properly choosing $q$, $m_{3/2}$ can be ultra-light and $M_{i}$ at the TeV scale.

Another way to disentangle the gaugino and gravitino masses and allow for low scale of SUSY breaking is 
also to consider Dirac mass terms for the gauginos \cite{Gherghetta:2011na,Goodsell:2014dia}. 
In particular just taking simply a Dirac gluino can relax strongly the LHC bounds on the neutralino 
sector, allowing for low Majorana neutralino masses, by avoiding universality between gauginos without
enhancing effects from the RGEs~\cite{Busbridge:2014sha}.

Within extra-dimensional models, even more possibilities arise. On one hand, many very light (pseudo)goldstino 
states could remain physical and play a similar role to the gravitino \cite{Benakli:2007zza} and on
the other the connections between superpartner masses become more involved.
A very interesting class of models was proposed recently in~\cite{Russell} and has the gravitino as a bulk 
field in a $(4+d)$-dimensional theory. The remaining particles in the MSSM mass spectrum are localized on a 
4-dimensional brane. Supersymmetry breaking occurs in a $(4+d')$-brane, with $d'\le d$, via an $F$-term 
denoted here by $F_{4+d'}$. After supersymmetry breaking, the MSSM particles acquire masses of the order 
of $M\sim 1/R\sim 1$~TeV, where $R$ is the size of the extra-dimensions. The gravitino KK tower instead 
starts at 
\begin{equation}
m_{3/2} \sim \frac{F_{4+d'}}{\sqrt{M_{*}^{2+d}/M^{d-d'}}}\,,
\end{equation}
which can be arbitrarily small, properly choosing $F_{4+d'}$, $d$, $d'$ and $M_{*}$, where $M_{*}$ is the 
fundamental gravitational scale. Notably, decays of the lightest particle in the MSSM mass spectrum into 
slightly lighter KK gravitinos are favored in this model. From the LHC perspective this model can hence 
resemble a 4-dimensional supersymmetric theory with a compressed mass spectrum.

\section{Cross-sections for gravitino production and scattering}
\begin{figure}[t]
\begin{center}
\includegraphics[width=0.45\textwidth]{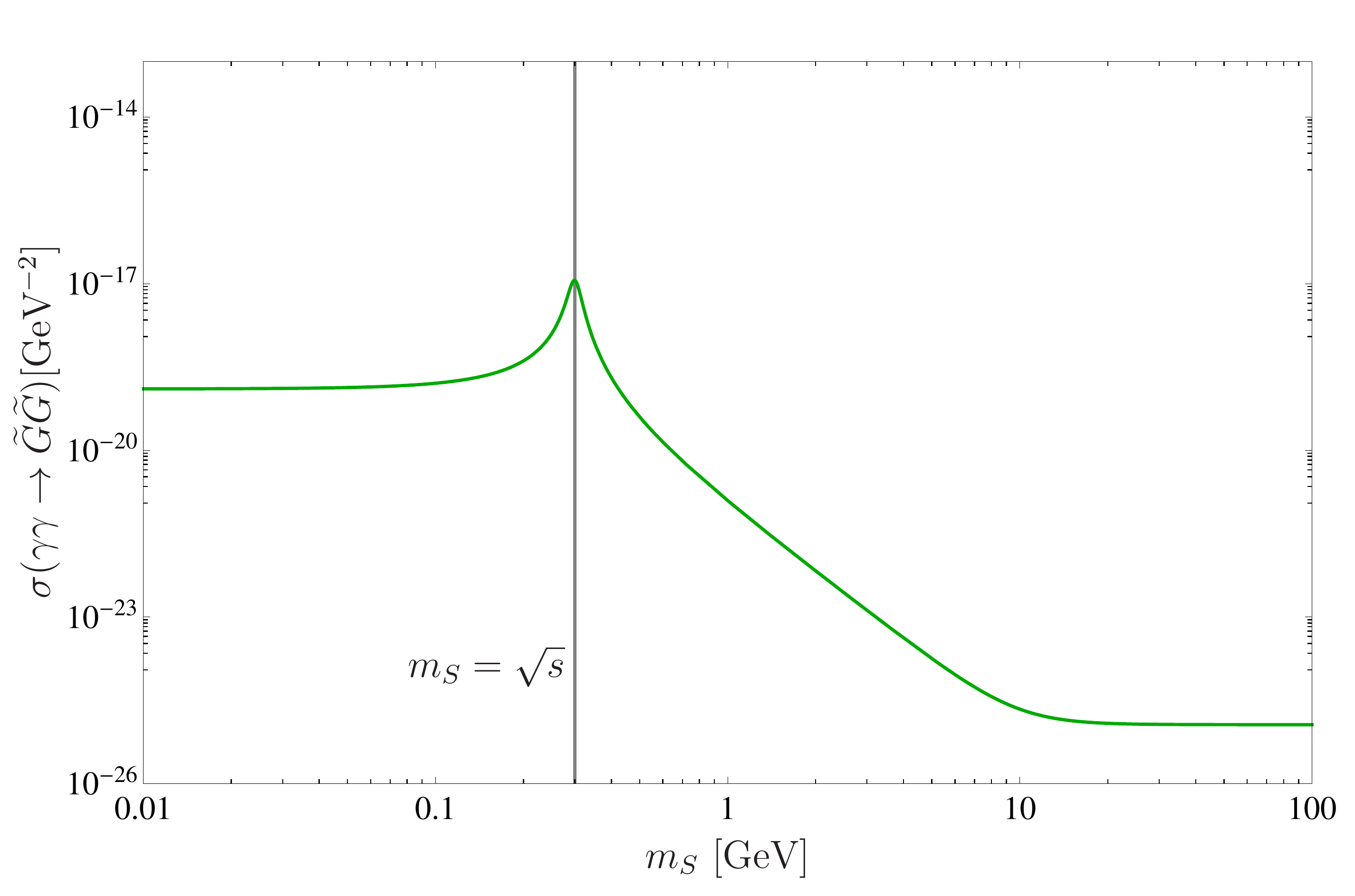}
\end{center}
\caption{The cross-section $\sigma(\gamma\gamma\rightarrow \widetilde{G}\widetilde{G})$ as a function of the sgoldstino mass $m_{S} = m_{P}$. Here we assume
$\Gamma_S=\Gamma_P=10^{-2}\, \text{GeV}$, $m_{\tilde{\gamma}}=$ 100\, GeV, $\xi \equiv c\cdot d = 10^{16}$ and $s=36 T_{\text{SN}}^2$ with $T_{\text{SN}}=50$ MeV.}
\label{fig:csplot}
\end{figure}

Given the general lagrangian in Eqs.~(\ref{eq:modelL}) and (\ref{eq:lsgoldstino}), we can then compute the 
relevant two-body processes. We discuss then possible effects of introducting R-parity breaking
in the model and find that those effects are negligible in our range of parameters.

\subsection{$\gamma\gamma\rightarrow \widetilde{G}\widetilde{G}$ cross-section and luminosity}
\label{ss:luminosity}
\begin{fmffile}{./feynman_diagrams}	
\fmfcmd{
				style_def gaugino expr p =
					shrink (1.5);			
						cdraw(photon p);
					endshrink;
					draw_plain p;
				enddef;
				style_def gravitino expr q =
					draw_double q;
				enddef;
				style_def spinor expr q =
					draw_plain q;
					shrink (0.7);			
						cfill(arrow q);
					endshrink;
					enddef;
				style_def matter expr q =
					draw_dashes q;
					shrink (0.7);			
						cfill(arrow q);
				endshrink;
					enddef;
					style_def fermionflow expr q =
					draw_plain q;
					shrink (0.6);			
						cfill(arrow q);
					endshrink;
				enddef;}
The dominant gravitino production contribution comes from photon-photon collisions \citep{grifols1996}. The expression for the corresponding luminosity can be found in app. \ref{app:luminosity}. The next step is to calculate $ \sigma(\gamma\gamma\longrightarrow \widetilde{G}\widetilde{G})$\footnote{We handle Majorana spinors by using the method of a continuous fermion flow \cite{denner1992}.}. The contributing diagrams are given by\vspace{0.5cm}
\begin{widetext}
\begin{align}
i\mathcal{M}&= \parbox{0.2 \textwidth}{
    \begin{fmfgraph*}(75,95)
    	\fmfpen{0.5 thin}
    	\fmfleft{i1,i2}
    	\fmfright{o1,o2}
    	\fmfv{label=$\alpha$,label.dist=1}{i2}
    	\fmfv{label=$\beta$,label.dist=1}{i1}
		\fmfv{label=$\nu$,label.dist=1}{o1}
		\fmfv{label=$\mu$,label.dist=1}{o2}
    	\fmf{photon,label=$p_2$,label.side=left,label.dist=1}{i1,v1}
    	\fmf{photon,label=$p_1$,label.side=right,label.dist=1}{i2,v2}
    	\fmf{gravitino,label=$k_2$,label.side=left,label.dist=1}{v2,o2}
    	\fmf{gravitino,label=$k_1$,label.side=left,label.dist=1}{o1,v1}
    	\fmf{gaugino,tension=0.8}{v1,v2}
    	\fmffreeze
    	\fmfforce{0.85w,0.9h}{r3}
    	\fmfforce{0.85w,0.1h}{r4}
    	\fmf{fermionflow,left=0.5,width=0.4}{r4,r3}
    \end{fmfgraph*}} + 
    \parbox{0.2 \textwidth}{
		 \begin{fmfgraph*}(75,95)
    	\fmfpen{0.5 thin}
    	\fmfleft{i1,i2}
    	\fmfright{o1,o2}
    	\fmfv{label=$\beta$,label.dist=1}{i2}
    	\fmfv{label=$\alpha$,label.dist=1}{i1}
		\fmfv{label=$\nu$,label.dist=1}{o1}
		\fmfv{label=$\mu$,label.dist=1}{o2}
    	\fmf{photon,label=$p_1$,label.side=left,label.dist=1}{i1,v1}
    	\fmf{photon,label=$p_2$,label.side=right,label.dist=1}{i2,v2}
    	\fmf{gravitino,label=$k_2$,label.side=left,label.dist=1}{v2,o2}
    	\fmf{gravitino,label=$k_1$,label.side=left,label.dist=1}{o1,v1}
    	\fmf{gaugino,tension=0.8}{v1,v2}
    	\fmffreeze
    	\fmfforce{0.85w,0.9h}{r3}
    	\fmfforce{0.85w,0.1h}{r4}
    	\fmf{fermionflow,left=0.5,width=0.4}{r4,r3}
    \end{fmfgraph*}
   	} +
   	\parbox{0.25 \textwidth}{
   		\begin{fmfgraph*}(95,75)
    	 \fmfpen{0.5 thin}
    	  \fmfleft{i1,i2}
    	  \fmfright{o1,o2}
 		\fmfv{label=$\alpha$,label.dist=1}{i2}
    	\fmfv{label=$\beta$,label.dist=1}{i1}
		\fmfv{label=$\nu$,label.dist=1}{o1}
		\fmfv{label=$\mu$,label.dist=1}{o2}
    	  \fmf{dbl_wiggly,tension=0.5}{v1,v2}
    	  \fmf{photon,label=$p_1$,label.dist=1,label.side=right}{i2,v1}
    	  \fmf{photon,label=$p_2$,label.dist=1,label.side=left}{i1,v1}
    	  \fmf{gravitino,label=$k_2$,label.dist=1,label.side=left}{v2,o2}
   		  \fmf{gravitino,label=$k_1$,label.dist=1,label.side=left}{o1,v2}
   		  \fmffreeze
 		\fmfforce{0.95w,0.9h}{r3}
    	\fmfforce{0.95w,0.1h}{r4}
    	\fmf{fermionflow,left=0.3,width=0.4}{r4,r3}
    \end{fmfgraph*} 
    }
   \nonumber \\[1cm]
&\qquad \qquad\qquad  + \parbox{0.25 \textwidth}{
     \begin{fmfgraph*}(95,75)
     \fmfpen{0.5 thin}
      \fmfleft{i1,i2}
      \fmfright{o1,o2}
 		\fmfv{label=$\alpha$,label.dist=1}{i2}
    	\fmfv{label=$\beta$,label.dist=1}{i1}
		\fmfv{label=$\nu$,label.dist=1}{o1}
		\fmfv{label=$\mu$,label.dist=1}{o2}
      \fmf{dashes,tension=0.5}{v1,v2}
    	  \fmf{photon,label=$p_1$,label.dist=1,label.side=right}{i2,v1}
    	  \fmf{photon,label=$p_2$,label.dist=1,label.side=left}{i1,v1}
    	  \fmf{gravitino,label=$k_2$,label.dist=1,label.side=left}{v2,o2}
   		  \fmf{gravitino,label=$k_1$,label.dist=1,label.side=left}{o1,v2}
      \fmffreeze
 		\fmfforce{0.95w,0.9h}{r3}
    	\fmfforce{0.95w,0.1h}{r4}
    	\fmf{fermionflow,left=0.3,width=0.4}{r4,r3}
    \end{fmfgraph*} 
    }+ \parbox{0.25 \textwidth}{
     \begin{fmfgraph*}(95,75)
     	\fmfpen{0.5 thin}
     	 \fmfleft{i1,i2}
     	 \fmfright{o1,o2}
 		\fmfv{label=$\alpha$,label.dist=1}{i2}
    	\fmfv{label=$\beta$,label.dist=1}{i1}
		\fmfv{label=$\nu$,label.dist=1}{o1}
		\fmfv{label=$\mu$,label.dist=1}{o2}
     	 \fmf{dots,tension=0.5}{v1,v2}
    	  \fmf{photon,label=$p_1$,label.dist=1,label.side=right}{i2,v1}
    	  \fmf{photon,label=$p_2$,label.dist=1,label.side=left}{i1,v1}
    	  \fmf{gravitino,label=$k_2$,label.dist=1,label.side=left}{v2,o2}
   		  \fmf{gravitino,label=$k_1$,label.dist=1,label.side=left}{o1,v2}
     	 \fmffreeze
 		\fmfforce{0.95w,0.9h}{r3}
    	\fmfforce{0.95w,0.1h}{r4}
    	\fmf{fermionflow,left=0.3,width=0.4}{r4,r3}
    \end{fmfgraph*} 
    }\label{eq:diagrams}\, .
\end{align}
\end{widetext}
This process has been calculated in \citep{bhattacharya1988,gherghetta1997} in the limit  of very small scalar and pseudo-scalar masses. Contrary to these studies, here we do not make any assumption regarding the sgoldstino masses, and assert no relation between the couplings $c$ and $d$. Accordingly, for the amplitude of this process we obtain,
\begin{align}
i\mathcal{M} &=i\mathcal{M}_{\text{Photino}}+i\mathcal{M}_{\text{Graviton}}+i\mathcal{M}_{\text{Scalar}} \nonumber\\ &+i\mathcal{M}_{\text{Pseudoscalar}}\, ,\label{eq:amplitude}
\end{align}
where the four invariant amplitudes $\mathcal{M}_{\text{Photino}}$, $\mathcal{M}_{\text{Graviton}}$, $\mathcal{M}_{\text{Scalar}}$ and $\mathcal{M}_{\text{Pseudoscalar}}$ are associated with the exchange of a photino, assumed here to be a mass-eigenstate, a graviton, a scalar and a pseudo-scalar, respectively. They are listed in the Appendix \ref{app:amplitudes}. 
For the calculation of $\overline{|\mathcal{M}|^2}$ and subsequently of the total cross-section we use the Mathematica package FeynCalc \citep{mertig1990}. 
We obtain the total cross-section, at leading order\footnote{We keep also terms $\mathcal{O}(m_{3/2}^{-2})$ which may be enhanced by the couplings $\propto m_{3/2}^{-1}$.}  $ 1/m_{3/2}^4 $,
\begin{eqnarray}
\sigma(\gamma\gamma\rightarrow \widetilde{G}\widetilde{G})
&=&\frac{s^3}{5760 \pi  m_{3/2}^4 M_P^4}\Bigg[1 + \frac{8}{3} \frac{m_{3/2}^2}{s}+5 \xi^2 \frac{m_{3/2}^2}{s} \nonumber\\  
&\times&  \left(\frac{s^2}{\left(s-m_P^2\right){}^2+\Gamma_P^2 \,m_P^2} \right. \nonumber\\ 
&+& \left.  \frac{s^2}{\left(s-m_S^2\right){}^2 +\Gamma_S^2 \,m_S^2}\right) \Bigg] + \mathcal{O}(\sqrt{x})\, , \label{eq:productioncs}
\end{eqnarray}
where $x\equiv s/m_{\tilde{\gamma}}^2\ll 1$, with $m_{\tilde{\gamma}}$ denoting the photino mass. Note that the photino t- and u-channel 
contribute to the cross-section only at the next order in $\sqrt{x}$ and therefore the leading order result is independent of the exact value of the photino mass and 
the possible corrections coming by considering the full neutralino mass matrix instead than the simple photino. 
The parameters $\Gamma_S$ and $\Gamma_P$, with dimension GeV, describe the model dependent decay widths of the scalar and pseudo-scalar $S, P$, respectively. 
Note that due to the non-renormalizable nature of the interaction, the cross-section grows with energy as $E^4 $ and therefore the best constraints are
obtained for the highest possible energy.
The cross-section is depicted in fig.~\ref{fig:csplot} as a function of $m_{S}=m_P$ for a typical average value of the energy in a SN core $s = (300 \mbox{MeV})^2$. 

The two limiting cases of very light or very heavy sgoldstinos, corresponding to the plateaus in fig.~\ref{fig:csplot}, are simply given by
\begin{align}
\lim_{m_S,m_P\rightarrow 0}\sigma(\gamma\gamma\rightarrow \widetilde{G}\widetilde{G}) &\approx \frac{s^3}{5760 \pi  m_{3/2}^4 M_P^4}\;  \nonumber\\
&\times \left(1+ \left(10 \xi^2 + \frac{8}{5}\right) \frac{m_{3/2}^2}{s}\right)\, ,\label{eq:ms0}\\
\lim_{m_S,m_P\rightarrow \infty}\sigma(\gamma\gamma\rightarrow \widetilde{G}\widetilde{G}) &\approx \frac{s^3}{5760 \pi  m_{3/2}^4 M_P^4}\, ,\label{eq:msinf}
\end{align}
so that we can nearly recover the limit of heavy sgoldstinos also by setting $\xi = 0$. For $\xi=\frac{m_{\tilde{\gamma}}}{m_{3/2}}$ the results are in agreement with \cite{bhattacharya1988,gherghetta1997,Brignole:1996fn}. The maximal value of the cross-section is obtained at the resonance and depends on the value of the widths $\Gamma_S, \Gamma_P $. So we expect the supernova environment to be most sensitive to sgoldstino with masses in the 100 MeV range. 
We substitute the cross-section in Eq.~\eqref{eq:productioncs} into Eq.~\eqref{eq:ggluminosity1} and we obtain the luminosity $L$ via integration.

Note that Eq.~(\ref{eq:ms0}) agrees with the results of~\cite{grifols1996} and can be used to obtain an upper limit on $m_{3/2}$ straightforwardly, without relying on the complex statistical analysis of Sec.~\ref{sec:statistics}. Using Eq.~\eqref{eq:ggluminosity1} as explained above, we obtain $m_{3/2}< 1.5 \times 10^{-5}$~eV, for $m_{\tilde{\gamma}}=100$~GeV. This result is comparable with the upper limit in~\cite{grifols1996}, though the two expressions differ because of an extra factor of 4 in our expression for $L$, and discrepancies in the numerical evaluation of the relevant formula. This finding will be refined within our Likelihood analysis.

\subsection{$\gamma\widetilde{G}\rightarrow \gamma\widetilde{G}$ cross-section and mean-free-path}

We have also to consider the mean-free path of the gravitino to compare with Eq.~\ref{eq:mfpconstraint}.
The dominant gravitino scattering process in the supernova core is $\gamma\widetilde{G}\rightarrow \gamma\widetilde{G}$, connected to the
production process by crossing symmetry. In fact the diagrams contributing to this scattering process are 
\begin{align}
i\mathcal{M}&= \parbox{0.1 \textwidth}{
    \begin{fmfgraph*}(50,50)
    	\fmfpen{0.5 thin}
    	\fmfleft{i1,i2}
    	\fmfright{o1,o2}
    	\fmf{photon}{i1,v1}
    	\fmf{photon}{o1,v2}
    	\fmf{gravitino}{v2,o2}
    	\fmf{gravitino}{i2,v1}
    	\fmf{gaugino,tension=0.8}{v1,v2}
    \end{fmfgraph*}} + 
    \parbox{0.1 \textwidth}{
		\begin{fmfgraph*}(50,50)
    		\fmfpen{0.5 thin}
      		\fmfleft{i1,i2}
      		\fmfright{o1,o2}
      		\fmf{photon}{i1,v2}
      		\fmf{photon}{v1,o2}
      		\fmf{gravitino}{i2,v1}
      		\fmf{gravitino}{v2,o1}
      		\fmf{gaugino}{v1,v2}
   		 \end{fmfgraph*} 
   	} +
   	\parbox{0.1 \textwidth}{
   		\begin{fmfgraph*}(50,50)
    	 \fmfpen{0.5 thin}
    	  \fmfleft{i1,i2}
    	  \fmfright{o1,o2}
    	  \fmf{dbl_wiggly,tension=0.5}{v1,v2}
      \fmf{photon}{i1,v1}
      \fmf{photon}{v1,o1}
      \fmf{gravitino}{i2,v2}
      \fmf{gravitino}{v2,o2}
    \end{fmfgraph*} 
    } \nonumber\\ &  \nonumber\\ & + \parbox{0.1 \textwidth}{
     \begin{fmfgraph*}(50,50)
     \fmfpen{0.5 thin}
      \fmfleft{i1,i2}
      \fmfright{o1,o2}
      \fmf{dashes,tension=0.5}{v1,v2}
      \fmf{photon}{i1,v1}
      \fmf{photon}{v1,o1}
      \fmf{gravitino}{i2,v2}
      \fmf{gravitino}{v2,o2}
    \end{fmfgraph*} 
    }  + \parbox{0.1 \textwidth}{
     \begin{fmfgraph*}(50,50)
     	\fmfpen{0.5 thin}
     	 \fmfleft{i1,i2}
     	 \fmfright{o1,o2}
     	 \fmf{dots,tension=0.5}{v1,v2}
      \fmf{photon}{i1,v1}
      \fmf{photon}{v1,o1}
      \fmf{gravitino}{i2,v2}
      \fmf{gravitino}{v2,o2}
    \end{fmfgraph*} 
    }\label{eq:diagrams2}\, .
\end{align}
We calculate the cross-section in the same way as in Sec.~\ref{ss:luminosity} and obtain
\begin{eqnarray}
\sigma(\gamma\widetilde{G}\rightarrow \gamma\widetilde{G}) &=& \frac{s^3}{768 \pi m_{3/2}^4 M_P^4} \Bigg[ 1 - 16 \frac{m_{3/2}^2}{s}  \nonumber\\
&+& \frac{\xi^2 m_{3/2}^2}{s+m_P^2} \Bigg(\frac{1}{6} -\frac{5}{18} \frac{m_P^2}{s} + \frac{5}{9} \frac{m_P^4}{s^2} \nonumber\\  
&-& \frac{5}{3} \frac{m_P^6}{s^3} - \frac{10}{3} \frac{m_P^8}{s^4} 
\left(1-\log\left[1+\frac{s}{m_P^2}\right]\right) \nonumber\\
&+& \frac{10}{3} \frac{m_P^{10}}{s^5} \log\left[1+\frac{s}{m_P^2}\right] \Bigg) \nonumber\\
&-& \xi^2 \frac{m_{3/2}^2}{s} \Bigg(+\frac{2
   s}{3 (s+m_S^2)}-\frac{5}{6} + \frac{10}{9} \frac{m_S^2}{s} \nonumber\\
 &-& \frac{5}{3} \frac{m_S^4}{s^2} + \frac{10}{3}\frac{m_S^6}{s^3}   \nonumber\\
&+& \frac{10}{3} \frac{m_S^8}{s^4} \log\left[\frac{m_S^2}{s+m_S^2}\right]\Bigg)\Bigg] 
+ \mathcal{O}(\sqrt{x})\, . \label{eq:scatteringcs}
\end{eqnarray} 
Finally, the gravitino mean-free-path $\lambda_{\rm mfp}$ in the Supernova core is
\begin{align}
\lambda_{\rm mfp} \simeq \left( n_{\gamma}(T_{SN})\sigma(\gamma\widetilde{G}\longrightarrow \gamma\widetilde{G})\right)^{-1}\label{eq:mfp}
\end{align}
with the photon number density $n_{\gamma}(T_{SN})$ given by
\begin{align}
n_{\gamma}(T) = \frac{2\zeta(3)}{\pi^2}T^3\, .\label{eq:averageenergy}
\end{align}
The average energy of a photon is $\sim 3 T$, corresponding to $s\sim 36 T^2$. For the Supernova core temperature we assume $T_{\text{SN}}\simeq 50$~MeV~\cite{raffelt1996}. Substituting \eqref{eq:scatteringcs} and \eqref{eq:averageenergy} into \eqref{eq:mfp} we obtain the mean-free-path depending on the sgoldstino masses and couplings as well as on the Supernova core temperature.

\subsection{Implications from R-Parity Violation}

Up to now we have assumed R-parity to be conserved\footnote{For an extensive review on R-parity and its violation we recommend \cite{barbier2004}.}. 
This is of course well-motivated e.g. from the non-observation of proton decay. With conserved R-parity,  gravitinos, or any other sparticle for that matter, 
can be produced in pairs only. If instead R-parity is broken, other channels giving the production of single gravitinos and containing only one supergravity 
vertex and less suppression by $ M_P $ become possible. It is therefore important to check if such additional contributions are stronger than the RPC ones
and can modify the constraints on the model.

The most important channel in our context is $\gamma\gamma\longrightarrow \nu \widetilde{G}$, since the only neutral SM fermions, that can be
produced together with the gravitino, are neutrinos. For the purpose of Supernova coupling, both gravitino and neutrino escape and therefore
this channel has the same effect as the RPC one.
We investigate whether this process is able to compete with $\gamma\gamma\longrightarrow\widetilde{G}\widetilde{G}$ for the case of leptonic 
trilinear and bilinear R-parity violations (RPV).\\

\subsubsection*{Trilinear RPV}
The possible trilinear RPV renormalizable couplings in the superpotential violating only the lepton number, are 
\begin{align}
W_{\slashed{R}_P} = \frac{1}{2}\lambda_{ijk} \epsilon_{ab} L^a_i L^b_j (E^c)_k + \lambda'_{ijk} \epsilon_{ab} L^a_i Q^b_j (D^c)_k\, , 
\label{eq:RPV}
\end{align}
where $i,j,k$ and $a,b$ are generation and $SU(2)$ doublet indices respectively. 
These couplings are only between the chiral multiplets and so do not involve photons. Nevertheless, at one-loop level they can
generate an effective photon-photino-neutrino vertex, that has been fully calculated in \citep{yamanaka2012}. 
In the loop we can have fermion-sfermion pairs of any particular flavor $j$, where $j=e,\mu,\tau,d,s,b$. 

In principle we also have other types of diagrams contributing to the same channel, e.g. box-diagram, but they include an
additional propagator and we therefore expect them to be more suppressed at low energy.

With the one-loop effective vertex the contributing diagrams are
\begin{widetext}
\begin{align}
i\mathcal{M}&=\sum_{i=1}^3\sum_j \left( \qquad \parbox{0.2 \textwidth}{
    \begin{fmfgraph*}(75,75)
    	\fmfpen{0.5 thin}
    	\fmfleft{i1,i2}
    	\fmfright{o1,o2}
    	\fmfv{label=$\alpha$,label.dist=1}{i2}
    	\fmfv{label=$\beta$,label.dist=1}{i1}
		\fmfv{label=$\mu$,label.dist=1}{o1}
		\fmfv{label=$\nu_i$,label.dist=1}{o2}
    	\fmf{photon,label=$p_2$,label.side=left,label.dist=1}{i1,v1}
    	\fmf{photon,label=$p_1$,label.side=right,label.dist=1}{i2,v2}
    	\fmf{spinor,label=$k_1$,label.side=left,label.dist=1}{v2,o2}
    	\fmf{gravitino,label=$k_2$,label.side=left,label.dist=1}{o1,v1}
    	\fmf{gaugino,tension=0.8}{v1,v2}
    	\fmffreeze
    	\fmfv{decor.shape=circle,decor.filled=shaded,decor.size=.15w,label=j,label.angle=-150}{v2}
    	\fmfforce{0.85w,0.9h}{r3}
    	\fmfforce{0.85w,0.1h}{r4}
    	\fmf{fermionflow,left=0.5,width=0.4}{r4,r3}
    \end{fmfgraph*}} + \qquad
    \parbox{0.2 \textwidth}{
		 \begin{fmfgraph*}(75,75)
    	\fmfpen{0.5 thin}
    	\fmfleft{i1,i2}
    	\fmfright{o1,o2}
    	\fmfv{label=$\beta$,label.dist=1}{i2}
    	\fmfv{label=$\alpha$,label.dist=1}{i1}
		\fmfv{label=$\mu$,label.dist=1}{o1}
		\fmfv{label=$\nu_i$,label.dist=1}{o2}
    	\fmf{photon,label=$p_1$,label.side=left,label.dist=1}{i1,v1}
    	\fmf{photon,label=$p_2$,label.side=right,label.dist=1}{i2,v2}
    	\fmf{spinor,label=$k_1$,label.side=left,label.dist=1}{v2,o2}
    	\fmf{gravitino,label=$k_2$,label.side=left,label.dist=1}{o1,v1}
    	\fmf{gaugino,tension=0.8}{v1,v2}
    	\fmffreeze
    \fmfv{decor.shape=circle,decor.filled=shaded,decor.size=.15w,label=j,label.angle=-150}{v2}
    	\fmfforce{0.85w,0.9h}{r3}
    	\fmfforce{0.85w,0.1h}{r4}
    	\fmf{fermionflow,left=0.5,width=0.4}{r4,r3}
    \end{fmfgraph*}
   	}
   	\right) \nonumber \\
   	&\equiv \sum_{i=1}^3 (i\mathcal{M}_{i})
\end{align}
\end{widetext}
Here we sum over all neutrino flavors $i$ in the final state. The blob $
\left(\;\,
\begin{fmfgraph*}(10,10)
    	\fmfpen{0.5 thin}
    \fmfforce{.5w,.3h}{i1}
\fmfv{decor.shape=circle,decor.filled=shaded,decor.size=w,label,label.angle=-180}{i1} 
\end{fmfgraph*}
\;\right)$
denotes an effective photon-photino-neutrino vertex, for
each fermionic flavour $j$. The explicit expression for the invariant amplitude is given in the Appendix \ref{app:amplitudes}.\\

Since we only need to perform a rough estimate in order to see the relevance of this channel, we do not
consider the full parameter dependence on the superparticle spectra and the full flavour structure of the
couplings. We take instead a maximal cross-section scenario with large flavour-democratic
couplings $\hat{\lambda}_{ijj} = \lambda \sim 0.1$, maximal sfermion mixing, $\sin \theta_{f_j}\cos \theta_{f_j} = \frac{1}{2}$,
vanishing CP phases and large mass hierarchy in the sfermion masses, i.e. we consider only the
contribution of the lighter mass eigenstate sfermion of any flavour, with a common mass $m^2_{\tilde{f}} \gg s $. 
In general the contributions of the two mass eigenstates of the same flavour tend to cancel, so this limit 
strongly enhances the scattering rate.

With these simplifications we obtain for the total cross-section
\begin{align}
\sum_i \sigma (\gamma\gamma\longrightarrow \nu_i \widetilde{G}) \approx 
\frac{49 \alpha^4 \lambda^2}{7680 \pi^3 \sin ^2\theta _W m_{3/2}^2  M_P^2} \frac{s^4}{m_{\tilde{\gamma}}^2 m_{\tilde{f}}^4}\; ,
\end{align}
where $\alpha $ is the QED coupling and $\theta_W $ the Weinberg angle. Note that indeed this channel is
suppressed only by two powers of $M_P$, but on the other hand also by the sixth power of the soft SUSY breaking masses.

We compare this result to the smallest possible cross-section for gravitino pair production i.e. \eqref{eq:msinf}, where we have no contributions 
from the sgoldstinos,
\begin{align*}
\frac{\sum_i \sigma (\gamma\gamma\longrightarrow \nu_i \widetilde{G})}{\sigma(\gamma\gamma\rightarrow \widetilde{G}\widetilde{G})} &\sim 10^{-6} 
\left( \frac{\lambda}{0.1}\right)^2 \left( \frac{m_{3/2}}{10^{-3}\text{eV}}\right)^2\, \nonumber\\ &\times \left( \frac{m_{\tilde\gamma}^2}{100\,\text{GeV}}\right)^{-2} 
\left( \frac{m_{\tilde{f}}}{100\,\text{GeV}}\right)^{-4}.
\end{align*}
It becomes clear that even under exaggerated choices of parameters single gravitino production due to trilinear RPV plays no role 
in our context. For gravitinos heavier than $ 1 $ eV the two processes do indeed become comparable, but the corresponding 
gravitino production would be too small to be relevant for SN cooling.

\subsubsection*{Bilinear RPV}

Alternatively we could also add bilinear R-parity violation (RPV) into our model by introducing the term
\begin{align}
W_{\slashed{R}_P} = \mu_i H_u\cdot L_i  \label{eq:bil}
\end{align}
to our superpotential. This term is also motivated by its ability to generate a hierarchical neutrino mass spectrum favored by observations. 

These models typically contain not only effective trilinear coupling as discussed above, but also neutralino-neutrino mixing since a sneutrino obtains 
in general a non-zero vacuum expectation value $\langle\tilde{\nu}\rangle$ during electroweak symmetry breaking \cite{ross1984}. 
In the context of supernova cooling bilinear RPV gives an additional contribution to the process 
$\gamma\gamma\rightarrow \widetilde{G}\nu$ with an intermediate neutrino. This channel is therefore not suppressed by the 
photino mass (even if it does influence the neutralino-sneutrino mixing). Contributing diagrams are
\begin{align}	
\parbox{0.25\textwidth}{
		\centering
		\begin{fmfgraph*}(70,100)
			\fmfpen{0.75 thin}
			\fmftop{t1,t2}
			\fmfbottom{b1,b2}
			\fmf{photon,label.dist=2,tension=1}{t1,v1}
			\fmf{photon,label.dist=2,tension=1,label.side=right}{b1,v3}
			\fmfv{label=$\alpha$,label.dist=1}{t1}
			\fmfv{label=$\beta$,label.dist=1}{b1}
			\fmfv{label=$\nu$}{b2}
			\fmfv{label=$\mu$,label.dist=1}{t2}
			\fmf{gaugino,tension=0.5,label.side=left,label=$\tilde{\chi}^0$}{v2,v1}
			\fmf{spinor,tension=0.5}{v2,v3}
			\fmf{gravitino,tension=1,label.dist=2}{v1,t2}
			\fmf{spinor,label.dist=2}{v3,b2}
			\fmffreeze
			\fmfleft{r1}
			\fmf{phantom}{r1,v4}
			\fmfv{decoration.shape=cross,label=$\langle\tilde{\nu}\rangle$,label.angle=180,decoration.size=8}{v4}
			\fmfset{dash_len}{2mm}
			\fmf{dashes,tension=1}{v2,v4}
			\fmfblob{.15w}{v3}
			\fmffreeze
			\fmfforce{1.0w,0.8h}{ff1}
			\fmfforce{1.0w,0.2h}{ff2}
			\fmf{fermionflow,right=0.7,width=0.4}{ff1,ff2}
		\end{fmfgraph*}	
	}+\parbox{0.25\textwidth}{
	\centering
	\begin{fmfgraph*}(70,100)
			\fmfpen{0.75 thin}
			\fmftop{t1,t2}
			\fmfbottom{b1,b2}
			\fmf{photon,tension=1}{t1,v1}
			\fmf{photon,tension=1,label.side=right}{b1,v3}
			\fmfv{label=$\beta$,label.dist=1}{t1}
			\fmfv{label=$\alpha$,label.dist=1}{b1}
			\fmfv{label=$\nu$}{b2}
			\fmfv{label=$\mu$,label.dist=1}{t2}
			\fmf{gaugino,tension=0.5,label.side=left,label=$\tilde{\chi}^0$}{v2,v1}
			\fmf{spinor,tension=0.5}{v2,v3}
			\fmf{gravitino,tension=1,label.dist=2}{v1,t2}
			\fmf{spinor,label.dist=2}{v3,b2}
			\fmffreeze
			\fmfleft{r1}
			\fmf{phantom}{r1,v4}
			\fmfv{decoration.shape=cross,label=$\langle\tilde{\nu}\rangle$,label.angle=180,decoration.size=8}{v4}
			\fmfset{dash_len}{2mm}
			\fmf{dashes,tension=1}{v2,v4}
			\fmfblob{.15w}{v3}
			\fmffreeze
			\fmfforce{1.0w,0.8h}{ff1}
			\fmfforce{1.0w,0.2h}{ff2}
			\fmf{fermionflow,right=0.7,width=0.4}{ff1,ff2}
		\end{fmfgraph*}	
			}
	\label{eq:diagramsrpv1}\, .
\end{align}
The blob in this diagram denotes this time the coupling of the neutrinos to the photon field. Since the neutrino is neutral this vertex is not present at tree level of course. 
However it can again be generated by radiative corrections, which lead to the real charge, magnetic dipole, electric dipole and anapole form factors of the neutrinos, which are strictly constrained by observations \cite{beringer2012}. It is now possible to determine whether the single gravitino production rate is of any relevance compared to the production of gravitino pairs from the same initial state\footnote{For the details on the calculation we refer to \cite{emken2013}.}.

Using the empirical bounds on the neutrino's electromagnetic properties and typical values for the RPV parameters, we find that the ratio of cross-sections is tiny even under the assumption of very heavy sgoldstinos, namely
\begin{align}
\frac{\sigma(\gamma\gamma\longrightarrow \tilde{G}\nu)}{\sigma(\gamma\gamma\rightarrow \tilde{G}\tilde{G})}& < 0.8\times 10^{-4} 
\left(\frac{m_{3/2}}{10^{-3}\;\text{eV}}\right)^2\left(\frac{\zeta}{10^{-5}}\right)^2\, ,\label{eq:vergleich1}
\end{align}
where $\zeta \equiv \frac{\langle\tilde{\nu}\rangle}{v}$ and $v$ is the SM Higgs VEV. Obviously the situation is the same as in the case of trilinear RPV, since
the parameter $ \zeta $ is constrained by the physical neutrino masses to be below $10^{-5} $ for neutralino masses at the electroweak scale.

In conclusion, in the area of parameter space relevant to SN cooling the production of gravitinos via photon collisions occurs almost exclusively in pairs. 
Neither modifications of the luminosity nor the mean-free-path can be of relevance making any further calculation or parameter scans including R-parity violations unnecessary. 

\end{fmffile}
\section{Model independent analysis}
In Sec.~\ref{sec:results}, we derive model independent limits on the mass of a light gravitino from Supernovae observations. To this aim, we compare the general 6-dimensional 
effective theory of the light gravitino interactions (Sec.~\ref{sec:theory}) to the bounds discussed in Sec.~\ref{sec:data} in a global statistical analysis. 
In Sec.~\ref{sec:statistics} we briefly describe the statistical methods used in this study. For a more extended introduction to statistical methods in particle physics and cosmology, 
we refer to the dedicated literature, e.g. Refs.~\cite{2008ConPh..49...71T,2011JHEP...06..042F,Conrad:2014nna}.

\subsection{Statistical framework}
\label{sec:statistics}
\begin{table}
    \begin{tabular}{lclc}
    Parameter         & Type & Prior range &  Prior type \\
    \hline
    $\log_{10} (m_{3/2}/{\rm GeV})$ & free parameter & $[-19,-12]$ & log-prior  \\
    $\log_{10} (m_{S}/{\rm GeV})$ & free parameter & $[-3,4]$ & log-prior  \\
    $\log_{10} (m_{P}/{\rm GeV})$ & free  parameter & $[-3,4]$ & log-prior  \\
    $\log_{10} (\xi)$ & free  parameter & $[0,18]$ & log-prior  \\
    $a$ & free   parameter & $[10^{-3},1]$ & linear prior  \\
    $b$ & free   parameter & $[10^{-3},1]$ & linear prior  \\
     \hline
    \end{tabular}
    \caption{List of model parameters. For each model parameter, this table shows the type of assumed prior PDF, and the corresponding prior range.}
    \label{tab:parameters}
\end{table}

We analyze the Supernovae data of Sec.~\ref{sec:data} combining Bayesian and frequentist statistics. In the Bayesian approach to data analysis, we calculate the posterior probability density function (PDF) of the model parameters $\mathbf{\Theta}=\theta_1,\dots\theta_n$, where $n$ is the dimension of the parameter space. In the present study $n=6$, and the six model parameters are listed in Tab.~\ref{tab:parameters}. The parameters $a\equiv \Gamma_S/m_S$ and $b\equiv \Gamma_P/m_P$ vary within a range where the narrow width approximation used in Eq.~(\ref{eq:productioncs}) can be applied. Note that the resonance in the cross-section results smoothed by the thermal average and therefore the final
results will not be strongly dependent on the sgoldstino widths.

In general, the posterior PDF, $\mathcal{P}(\mathbf{\Theta} | \mathbf{d})$, describes the observer's degree of belief on a certain set of model parameters, after having analyzed the available data, $\mathbf{d}$. It can be calculated applying Bayes' theorem to $\mathcal{L}(\mathbf{d}|\mathbf{\Theta})$, the Likelihood function of the data:
\begin{equation}
\mathcal{P}(\mathbf{\Theta} | \mathbf{d}) = \mathcal{L}(\mathbf{d} | \mathbf{\Theta}) \pi(\mathbf{\Theta})\mathcal{E}(\mathbf{d})^{-1} \,.
\label{eq:bayes}
\end{equation} 
In Eq.~(\ref{eq:bayes}), $\mathcal{E}(\mathbf{d})$ is the Bayesian evidence. It is independent of the model parameters, and it hence plays the role of a normalization constant in the present analysis. $\pi(\mathbf{\Theta})$ is the prior PDF. It contains the observer's knowledge of the model parameters, before having analyzed the available data. In the analyses we assume log-priors for all model parameters, except for $a$ and $b$. This assumption allows to effectively sample the posterior PDF varying the model parameters over several order of magnitudes. Tab.~\ref{tab:parameters} shows the range of variability that we assume for the different model parameters. The intervals chosen for $m_S$ and $m_P$ allow us to effectively explore the two extreme regimes $m_S,m_P\rightarrow 0$ and  $m_S,m_P\rightarrow \infty$, where, respectively, $m_{S,P}/T\ll 1$ and $m_{S,P}/T\gg 1$.

We express our results in terms of 1D and 2D marginal posterior PDFs. The 2D marginal posterior PDF of the model parameters $\theta_1$ and $\theta_2$, for instance,  is defined as follows
\begin{equation}
\mathcal{P}_{\rm marg}(\theta_1, \theta_2|\mathbf{d})  \propto \int d \theta_{3}\dots d \theta_{n}\,  \mathcal{P}(\mathbf{\Theta} | \mathbf{d})  \,.
\label{eq:marg}
\end{equation} 
Similar expressions hold for the other pairs of model parameters. From the 2D marginal posterior PDFs, we construct the associated 95\% credible regions. They contain the 95\% of the total posterior probability, and are such that $\mathcal{P}_{\rm marg}$ at any point inside the region is larger than at any point outside the region. 

When the prior PDF contains more information than the Likelihood function, the support of the posterior PDF extends over a large volume in parameter space. In this case, the integral in Eq.~(\ref{eq:marg}) mostly depends on the integration volume, rather then on the information contained in the Likelihood function. A similar dependence on the integration volume in Eq~(\ref{eq:marg}) is called a ``volume effect''.

In the frequentist approach, from the Supernovae bounds we extract the 1D and 2D profile likelihoods of the model parameters. 
The 2D profile likelihood of the model parameters $\theta_1$ and $\theta_2$, for instance, is given by
\begin{equation}
\mathcal{L}_{\rm prof}(\mathbf{d}|\theta_1, \theta_2) \propto \max_{\theta_3\dots\theta_n} \mathcal{L}(\mathbf{d} | \mathbf{\Theta})  \,,
\label{eq:prof_likelihood}
\end{equation} 
and analogously for the other pairs of model parameters. We use the profile likelihood to construct approximate frequentist confidence intervals from an effective chi-square defined as $\Delta \chi^2_{\rm eff}\equiv-2 \ln \mathcal{L}_{\rm prof}/ \mathcal{L}_{\rm max}$, where $\mathcal{L}_{\rm max}$ is the absolute maximum of the Likelihood function. 

The Bayesian approach and the frequentist approach are complementary. The former allows to obtain limits on the gravitino mass with a relatively small number of Likelihood evaluations. The latter is computationally demanding, but it does not crucially depend on the choice of the prior PDF, and it is not affected by volume effects. 

A simple Likelihood function which can be used in presence of an upper limit on the gravitino luminosity is given by the following expression 
\begin{equation}
\mathcal{L}(\mathbf{d} | \mathbf{\Theta}) = \left\{
\begin{array}{lc}
1 &  \qquad\qquad {\rm for}~L \le L_{\rm exp} \\
& \\
\exp \left[-\frac{(L - L_{\rm{exp}})^2}{2\sigma_{\rm exp}^2}\right] & \qquad\qquad  {\rm otherwise},
\end{array}
\right. 
\label{eq:Like1}
\end{equation} 
where $\sigma_{\rm exp}$ is an estimate of the experimental error on $L_{\rm exp}$. Following \cite{Austri:2006pe}, in the analyses we adopt a refined version of Eq.~(\ref{eq:Like1}) which includes theoretical and experimental errors for the gravitino luminosity, denoted here by $\tau$ and $\sigma_{\rm exp}$, respectively. The Likelihood function used in the calculations is hence given by
\begin{align}
\mathcal{L}(\mathbf{d} | \mathbf{\Theta}) &= \frac{\sigma_{\rm exp}}{\sqrt{\sigma_{\rm exp}^2+\tau^2}} \exp\left[ - \frac{(L_{\rm exp} - L)^2}{2(\sigma_{\rm exp}^2+\tau^2)}\right] \nonumber\\ &\times \left[1-Z(t_{\rm exp})\right] + Z\left( \frac{L-L_{\rm exp}}{\tau} \right)
\label{eq:Like2}
\end{align}
for $\lambda_{\rm mfp}\ge\lambda_{\rm exp}=0.3$~m, and $\mathcal{L}(\mathbf{d} | \mathbf{\Theta}) = 1$ otherwise, when the Supernovae constrain does not apply. The function $Z$ in Eq.~(\ref{eq:Like2}) is defined as follows
\begin{equation}
Z(y) \equiv \frac{1}{\sqrt{2\pi}}\int_y^{\infty} dx\,\exp\left( -x^2/2 \right)
\end{equation}
and 
\begin{equation}
t_{\rm exp} \equiv \frac{\sigma_{\rm exp}}{\tau} \frac{L-L_{\rm exp}}{\sqrt{\sigma_{\rm exp}^2+\tau^2}} \,. 
\end{equation}
In the calculations we assume $\sigma_{\rm exp}/L_{\rm exp}$  and $\tau/L_{\rm exp}$ equal to 0.05. Finally, to sample the Likelihood function we use the {\sffamily Multinest} program~\cite{Feroz:2008xx,Feroz:2007kg,Feroz:2013hea}, with parameters set to: $n_{\rm live}=20000$ and tol=$10^{-4}$. We use our own routines to calculate the cross-sections and the mean-free-path. Figures are produced using the programs {\sffamily Getplots}~\cite{Austri:2006pe} and {\sffamily Matlab}. 

\subsection{Limits on the gravitino mass}
\label{sec:results}
\begin{figure*}[t]
\begin{center}
\includegraphics[width=0.8\textwidth]{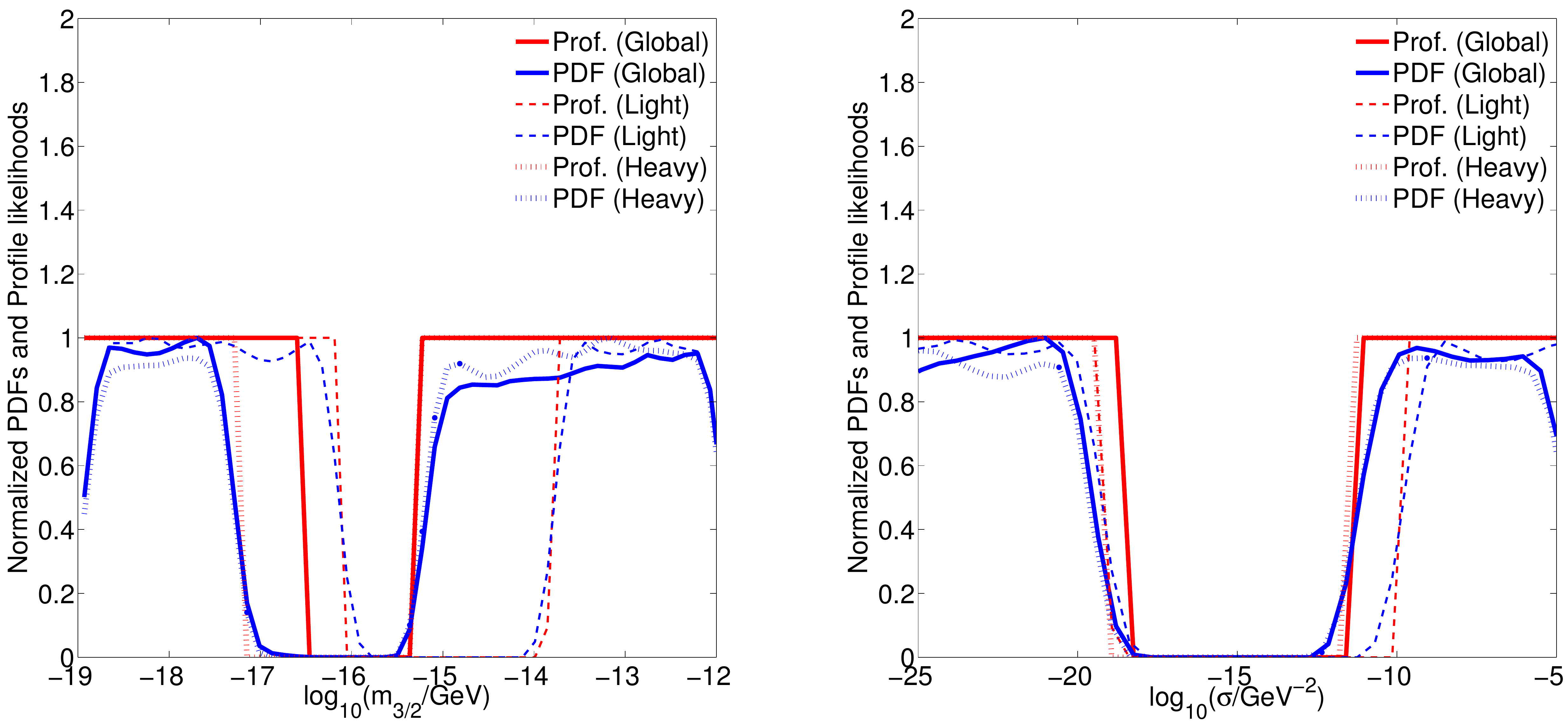}
\end{center}
\caption{{\it Left panel.} Limits on the gravitino mass found from Supernovae bounds: (1) A model with heavy scalar and pseudo-scalar fields, $a=b=0$, and $\xi=100~{\rm GeV})/m_{3/2}$ (dotted lines). (2) A model with light scalar and pseudo-scalar fields, $a=b=0$, and $\xi=(100~{\rm GeV})/m_{3/2}$ (dashed lines). (3) The general 6-dimensional effective theory of the light gravitino interactions. Red lines denote 1D profile likelihoods, blue lines correspond to 1D marginal posteriors PDFs. Analyzing the general 6-dimensional effective theory of the light gravitino interactions, we find the model independent exclusion limits on the gravitino mass $-16.65<\log_{10}(m_{3/2}/{\rm GeV})<-15.27$ at the 95\% confidence level (corresponding to $2.3\times10^{-8} <m_{3/2}/{\rm eV}<5.4\times 10^{-7} $). {\it Right panel.} As for the left panel, but for the cross-section $\sigma(\gamma\gamma\rightarrow\widetilde{G}\widetilde{G})$ evaluated at the reference centre-of-mass energy $s=36\,T^2$.}
\label{fig:limits}
\end{figure*}

We now describe the results obtained comparing the effective theory of the light gravitino interactions to the Supernovae bounds. 
In the analyses we use the statistical methods introduced in Sec.~\ref{sec:statistics}. 
Note that on general grounds we expect to exclude a window of gravitino masses, limited on the lower side by the requirement to have a sufficiently long
mean-free path in the SN core and on the upper side by the SN luminosity bound.

\begin{figure*}[t]
\begin{center}
\includegraphics[width=0.8\textwidth]{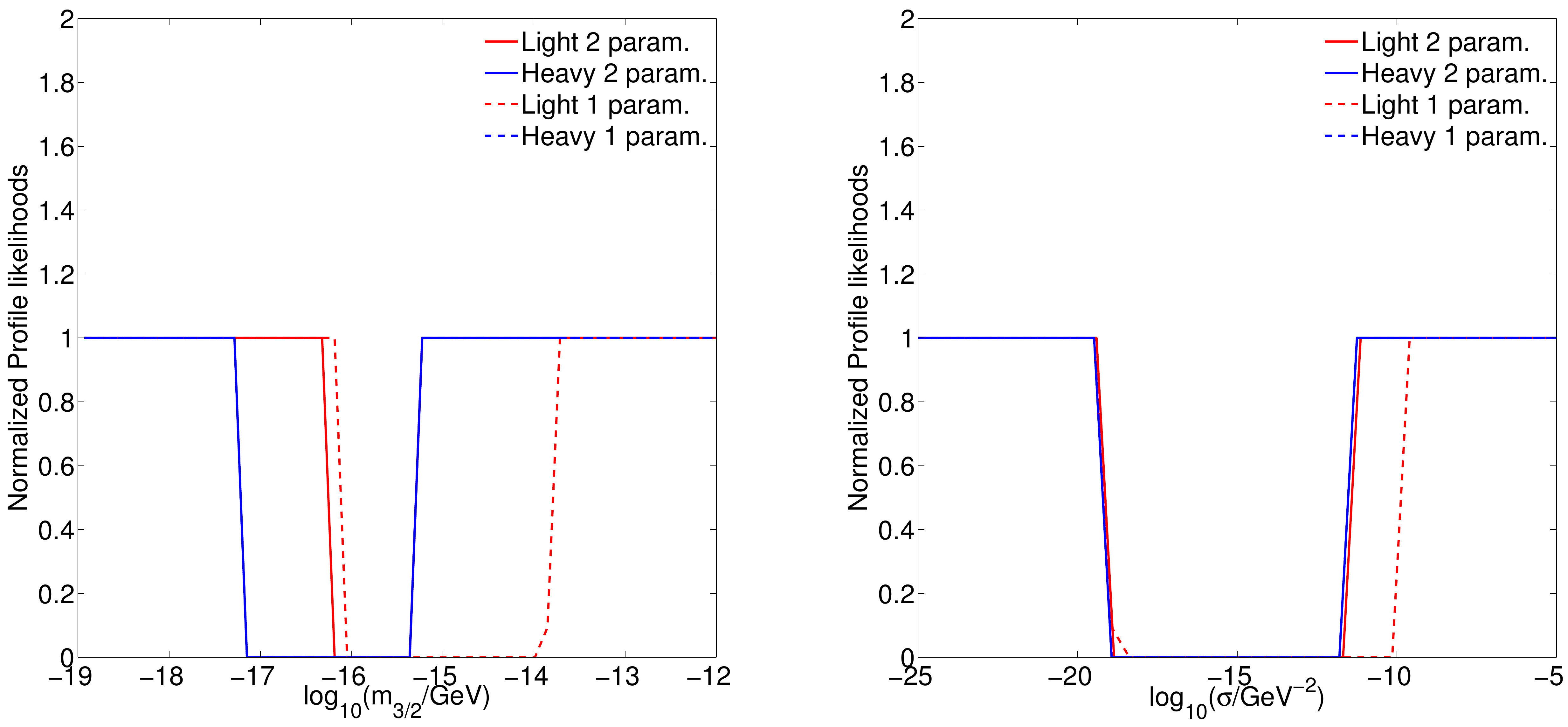}
\end{center} 
\caption{{\it Left panel.} As for Fig.~\ref{fig:limits} (profile likelihoods only), but for 4 different models: (1) A model with light scalar and pseudo-scalar fields, $a=b=0$, and $\xi$ and $m_{3/2}$ as free parameters (red solid line). (2) A model with heavy scalar and pseudo-scalar fields, $a=b=0$, and $\xi$ and $m_{3/2}$ as free parameters (blue solid line). (3) Same as model 2 in Fig.~\ref{fig:limits} (red dashed line). (4) Same as model 1 in Fig.~\ref{fig:limits} (blue dashed line). Model 2 and model 4 in this figure are characterized by identical curves, since the gravitino production cross-section is independent of $\xi$ in the heavy mass limit. Interestingly, Model 1 gives results similar to the full 6 dimensional model in Fig.~\ref{fig:limits}. {\it Right panel.} As for the left panel, but now for the cross-section $\sigma(\gamma\gamma\rightarrow\widetilde{G}\widetilde{G})$ at $s=36\,T^2$.}
\label{fig:limits2}
\end{figure*}
The left panel in Fig.~\ref{fig:limits} shows the limits on the gravitino mass derived within three different realizations of the theory defined in Sec.~\ref{sec:theory}. 
Red lines refer to 1D profile likelihoods, whereas blue lines correspond to 1D marginal posterior PDFs. The red dashed line in the left panel of Fig.~\ref{fig:limits} is the 
1D profile likelihood of a model with light scalar and pseudo-scalar fields (i.e. $m_{S}\rightarrow0$ and $m_{P}\rightarrow0$) with fixed sgoldstino coupling
$\xi = 100 \mbox{GeV}/m_{3/2} $.  From the 1D profile likelihood of  this model, we are able to exclude the mass range $-16.19~<~\log_{10}(m_{3/2}/{\rm GeV})<-13.85$ at the 95\% confidence level (corresponding to $6.5\times10^{-8} <m_{3/2}/{\rm eV}<1.4\times 10^{-5} $). This mass window is comparable (up to the factor 4) to the old exclusion 
obtained in \cite{grifols1996} with the same assumption on the value of  $\xi$.
The blue dashed line in the left panel of Fig.~\ref{fig:limits} is the 1D marginal PDF derived from the same model. Compared to the 1D profile likelihood, 
it increases from its minimum value less steeply and it is not flat at the boundaries of the prior range. The latter effect is related to the prior PDF, which forces to zero 
the posterior PDF at the edge of the prior range.

The red dotted and blue dotted lines in the left panel of Fig.~\ref{fig:limits} are, respectively,  the 1D profile likelihood and the 1D marginal posterior PDF of a model with heavy scalar and pseudo-scalar fields (i.e. $m_{S}\rightarrow\infty$ and $m_{P}\rightarrow\infty$). From the 1D profile likelihood of this case, we find the exclusion limits $-17.28<\log_{10}(m_{3/2}/{\rm GeV})<-15.36$ at the 95\% confidence level (corresponding to $5.2\times10^{-9} <m_{3/2}/{\rm eV}<4.3\times 10^{-7} $). 
Comparing the 1D profile likelihood to the 1D marginal posterior PDF, we observe the same differences found in the case of light scalar and pseudo-scalar fields. 

The red solid and blue solid lines in the left panel of Fig.~\ref{fig:limits} are, respectively,  the 1D profile likelihood and the 1D marginal posterior PDF obtained fitting the full 6-dimensional effective theory of Sec.~\ref{sec:theory} to the Supernovae data, without any assumption on the sgoldstinos masses and coupling. The exclusion limits on the gravitino mass extracted from the 1D profile likelihood of this general model are one of the main results of this paper. Analyzing the general 6-dimensional effective theory of the light gravitino interactions, we find the following model independent exclusion limits for the gravitino mass: 
\begin{equation}
-16.65<\log_{10}(m_{3/2}/{\rm GeV})<-15.27
\label{eq:main_limit}
\end{equation}
at the 95\% confidence level (corresponding to $2.3\times10^{-8} <m_{3/2}/{\rm eV}<5.4\times 10^{-7} $). 
We see that the mass window excluded in a model-independent way is much narrower than the ranges for the two limiting cases, partially due to the resonant enhancement 
in the full cross-section. 
On the other hand these limits are very robust, since they take into account the uncertainties in the masses and couplings defining the model of Sec.~\ref{sec:theory} in a statistically rigorous manner. The 1D marginal posterior PDF for $m_{3/2}$, that we find marginalizing the full 6-dimensional posterior PDF over the 5 remaining model parameters, is significantly different from the 1D profile likelihood of the same 6-dimensional model. The differences between the two statistical indicators are induced by volume effects produced in the marginalization procedure, as we will see in detail below.

The right panel in Fig.~\ref{fig:limits} shows the 1D profile likelihoods and the 1D marginal posterior PDFs for the cross-section $\sigma(\gamma\gamma\rightarrow\widetilde{G}\widetilde{G})$ at $s = 36\, T^2$ in the three models discussed in the left panel of the same figure. From an analysis based on the general 6-dimensional effective theory of the light gravitino interactions, we extract the 95\% confidence level exclusion limits $-19.26<\log_{10}(\sigma/{\rm GeV}^{-2})<-11.85$, for $\sigma(\gamma\gamma\rightarrow\widetilde{G}\widetilde{G})$ at $s = 36\,T^2$. 

\begin{figure*}[t]
\begin{center}
\includegraphics[width=0.9\textwidth]{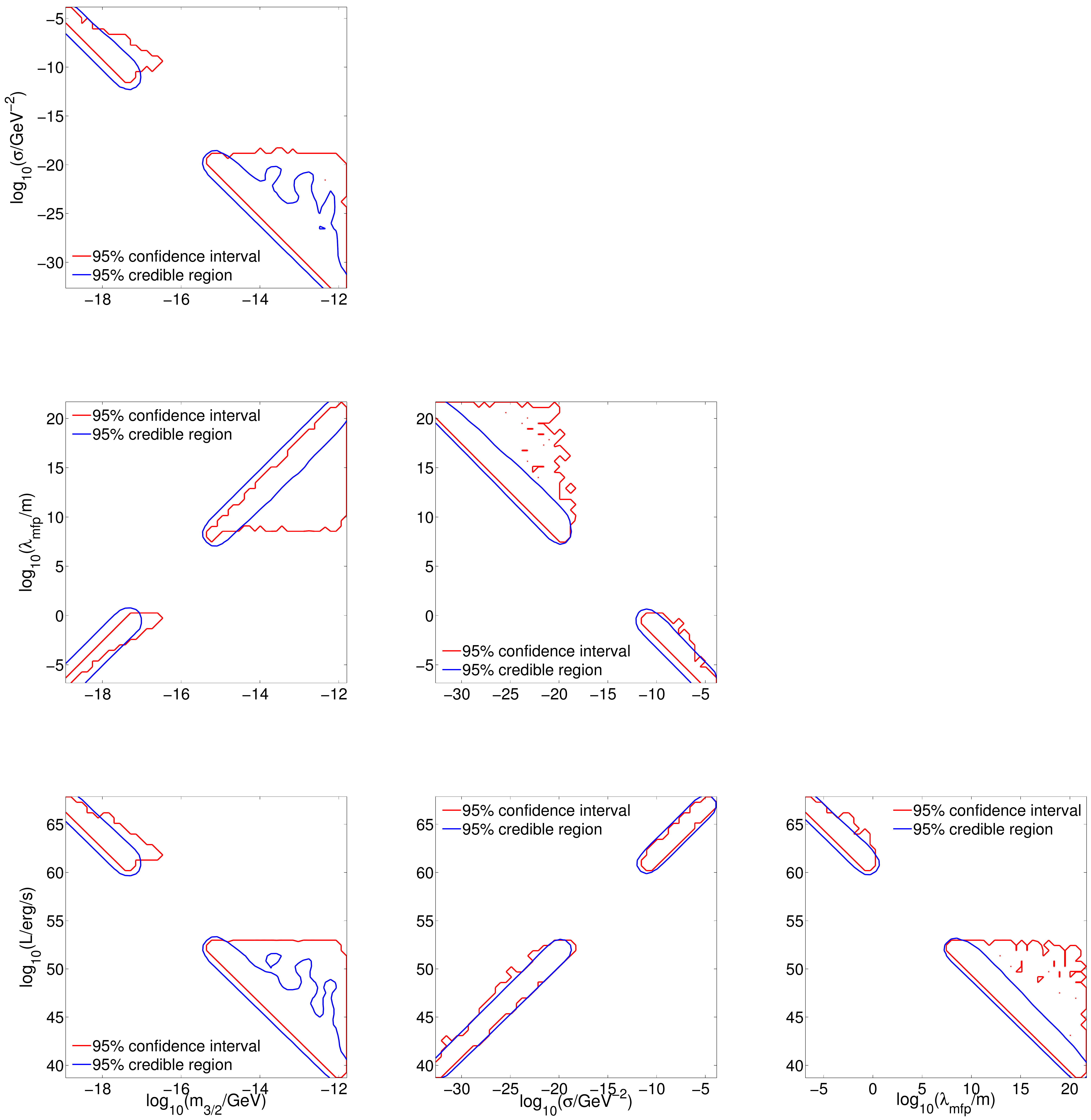}
\end{center}
\caption{2D 95\% confidence intervals (red contours) and 2D 95\% credible regions (blue contours) in the six planes $(m_{3/2},L)$,  $(m_{3/2},\lambda_{\rm mfp})$, $(m_{3/2},\sigma)$,  $(\sigma,L)$, $(\sigma,\lambda_{\rm mfp})$ and $(\lambda_{\rm mfp},L)$. The contours have been derived fitting the general 6-dimensional effective theory of Sec.~\ref{sec:theory} to the Supernovae data.}
\label{fig:correlations}
\end{figure*}
\begin{figure}[t]
\begin{center}
\includegraphics[width=0.35\textwidth]{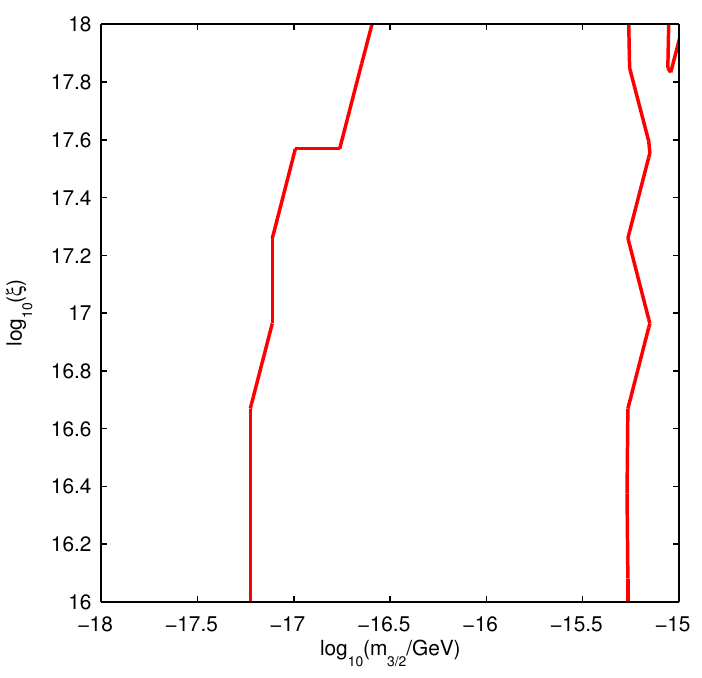}
\end{center}
\caption{2D 95\% confidence interval (red contour) in the $m_{3/2}-\xi$ plane. The parameters $m_{3/2}$ and $\xi$ are degenerate in the large $\xi$ limit. Inspection of Eqs.~(\ref{eq:productioncs}) and (\ref{eq:scatteringcs}) shows that only the ratio $\xi^2/m_{3/2}^2$ is constrained by observations in this limit.}
\label{fig:m32xi}
\end{figure}

We also study additional models with 2 free parameters. In Fig.~\ref{fig:limits2} we compare their 1D profile likelihoods. A first model is characterized by $m_{S}=m_{P}=0$, $a=b=0$, and $\xi$ and $m_{3/2}$ as free parameters (red solid line). A second model is defined by $m_{S,P}\rightarrow \infty$, $a=b=0$, and again $\xi$ and $m_{3/2}$ as free parameters (blue solid line). Comparing Figs.~\ref{fig:limits} and \ref{fig:limits2}, we see that the model with $\xi$ and $m_{3/2}$ as free parameters and $m_S,m_P\rightarrow 0$, and the full 6-dimensional model have similar 1D profile likelihoods, and confidence intervals, though the former are slightly narrower. For instance, within the 2-parameter model, we find the 95\% confidence level exclusion limits for $m_{3/2}$: $-16.32<\log_{10}(m_{3/2}/{\rm GeV})<-15.36$, corresponding to $4.8\times10^{-8} <m_{3/2}/{\rm eV}<4.3\times 10^{-7}$. Indeed the two parameters $\xi$ and $m_{3/2}$ capture the strongest dependence of the cross-section. 

Let us now go back to consider the general case with 6 free parameters. Fig.~\ref{fig:correlations} shows the 2D 95\% confidence intervals (red lines) and the 95\% credible regions (blue lines) in the six planes $(m_{3/2},L)$,  $(m_{3/2},\lambda_{\rm mfp})$, $(m_{3/2},\sigma)$,  $(\sigma,L)$, $(\sigma,\lambda_{\rm mfp})$ and $(\lambda_{\rm mfp},L)$, as obtained by comparing the general 6-dimensional theory of Sec.~\ref{sec:theory} to the Supernovae constraint. As in Fig.~\ref{fig:limits}, $\sigma$ is evaluated at the reference centre-of-mass energy $s = 36\,T^2$. The bottom-left panel in Fig.~\ref{fig:correlations} shows the results for the pair of model parameters $m_{3/2}$ and $L$. In this panel, the 2D 95\% confidence interval and the 95\% credible region separate in two disconnected areas. One at large $m_{3/2}$ generated by the Supernovae bound on the gravitino luminosity, and one at at small $m_{3/2}$, where $\lambda_{\rm mfp}<0.3$~m and the supernovae bound does not apply. 

Models with constrained $\xi \propto m_{3/2}^{-1} $, for finite sgoldstino masses, or large sgoldstino masses, predict $\log_{10}(L) = -4 \log_{10} (m_{3/2})+{\rm const.}$. The diagonal structure observed in the 
bottom-left panel of Fig.~\ref{fig:correlations} reflects this correlation pattern between $L$ and $m_{3/2}$. However, in the general 6-dimensional theory of Sec.~\ref{sec:theory}, the dependence of $L$ on $m_{3/2}$ is less obvious. For this reason we also observe ``out-of-diagonal structures'' in the $(m_{3/2},L)$ plane, corresponding to configurations with 
intermediate values of $m_{S}$ and $m_P$, and values of $\xi$ different from $m_{\tilde{\gamma}}/m_{3/2}$. The 2D credible region in the bottom-left panel of Fig.~\ref{fig:correlations} 
is concentrated along the diagonal of the plane ($m_{3/2}$,$L$). This result is related to the volume effect also appearing in Fig.~\ref{fig:limits}. In fact, marginalizing over the large 
volume in parameter space where $m_S$, $m_P$ and $\xi$ are large\footnote{$m_S$ and $m_P$ larger than the Supernovae core temperature, and $\xi$ very large.} leads to an 
artificial overweighting of configurations following the 
correlation pattern $\log_{10}(L) = -4 \log_{10} (m_{3/2})+{\rm const}$. Therefore, the Bayesian approach tends to favour configurations 
along the diagonal in the plane ($m_{3/2}$,$L$), because of spurious volume effects. The presence of out-of-diagonal structures in the 2D profile likelihood also induces the differences 
observed in Fig.~\ref{fig:limits} between the 1D profile likelihood and the 1D marginal posterior PDF of the general 6-dimensional model. We therefore conclude that only the frequentist 
approach produces robust and physically relevant results, since the Bayesian approach is affected by volume effects, when present Supernovae data are used.

The central-left and top-left panels in Fig.~\ref{fig:correlations} refer to the pairs of model parameters $(m_{3/2},\lambda_{\rm mfp})$ and $(m_{3/2},\sigma)$, respectively. The pair $(m_{3/2},\sigma)$ exhibits a statistical behavior essentially identical to the related pair $(m_{3/2},L)$. Also the results found for the pair $(m_{3/2},\lambda_{\rm mfp})$ admit an analogous interpretation, with one important difference, however. In the case of the pairs $(m_{3/2},\lambda_{\rm mfp})$, the two parameters are positively correlated and the correlation pattern expected for fixed $\xi \propto m_{3/2}^{-1} $ at small scalar masses is $\log_{10}(\lambda_{\rm mfp}) = 4 \log_{10} (m_{3/2})+{\rm const}$.

The remaining panels show the correlations between $L$, $\sigma$ and $\lambda_{\rm mfp}$. As expected, we find that $L$ and $\sigma$ are positively correlated, whereas $L$ and $\lambda_{\rm mfp}$, and $\sigma$ and $\lambda_{\rm mfp}$ are negatively correlated.

Finally, we briefly comment on possible degeneracies between pairs of model parameters. Inspection of Eqs.~(\ref{eq:productioncs}) and (\ref{eq:scatteringcs}) shows that only the ratio $\xi^2/m_{3/2}^2$ can be constrained by observations in the large $\xi$ limit. This degeneracy is captured by the numerical analysis, as Fig.~\ref{fig:m32xi} shows. Calculating confidence intervals, we do not observe strong degeneracies between other pairs of model parameters 
not even between $ \xi^2 $ and $ m_{S,P} $ in the heavy sgoldstino limits. 

\section{Conclusions} 

We have revisited Supernovae constraints on ultralight gravitinos in a more modern and model-independent way, 
extending them to general SUSY breaking models and including as well a discussion of
possible effects of lepton number breaking R-parity violation. For what regards the latter, we find that the
single gravitino production channels are always negligible compared to the R-parity conserving two
gravitino processes for the ultralight masses relevant for the Supernovae.

Our analysis was therefore focussed on the RPC general models, where we let the sgoldstino masses and couplings 
vary within reasonable expected ranges. We found that there are two very clearly distinguishable regimes
of light and heavy sgoldstinos, where light/heavy is meant as compared to the SN average thermal energy scale of $ ~ 100 $ MeV.
In the case of light sgoldstinos, the cross-section is larger and our frequentist analysis is in agreement with the old results 
and gives comparable limits, for sgoldstino couplings $\xi = 100~\mbox{GeV}/m_{3/2} $. In the heavy sgoldstino region, the cross-section
is slightly reduced and in order to compensate that, the excluded window is therefore slightly shifted to lower gravitino masses. 
Note that the two different excluded ranges do overlap and are given at the 95\% confidence level by
\begin{align}
& 6.5\times10^{-8}~{\rm eV} < m_{3/2} <1.4\times 10^{-5}~{\rm eV}\,, \nonumber\\ &  \mbox{for}\quad \xi = 100~\mbox{GeV}/m_{3/2} \quad m_{S,P} \rightarrow 0 \,; 
\nonumber\\
& 5.2\times10^{-9}~{\rm eV} <m_{3/2} <4.3\times 10^{-7}~{\rm eV}\,,  \nonumber\\ & \mbox{for}\quad \xi = 100~\mbox{GeV}/m_{3/2} \quad m_{S,P} \rightarrow\infty \,.
\nonumber
\end{align}
The bayesian analysis for both cases excludes very similar ranges of the gravitino mass, but is unfortunately limited 
by volume effects.

Considering the whole range of parameters and free sgoldstino masses, we are able to set a new model-independent limit 
on the gravitino mass as
\begin{equation}
2.3 \times 10^{-8}~{\rm eV} < m_{3/2} < 5.4 \times 10^{-7}~{\rm eV}
\label{eq:final}
\end{equation}
at the 95\% confidence level. We note that this excluded window corresponds more or less to the overlap
region of the two limiting cases and is therefore reduced to be quite narrow. We also mention that a 
simplified analysis based on the model parameters $m_{3/2}$ and $\xi$ only, produces exclusion limits 
similar to the model independent limit in Eq.~(\ref{eq:final}), though slightly narrower.

The gravitino mass window in Eq.~(\ref{eq:final}) is within the region ruled out by BBN in the limit of massless sgoldstinos never in thermal equilibrium. 
For instance, one can exclude $m_{3/2}\lesssim 10^{-6}~{\rm eV} (m_{\tilde{\gamma}}/100~{\rm GeV})^{1/2}$, demanding 
that the goldstino decoupling temperature is larger than about 100 MeV~\cite{gherghetta1997,grifols1996}. 
We however expect that for heavy sgoldstinos the same argument of~\cite{gherghetta1997,grifols1996} would produce weaker limits on $m_{3/2}$.
If also the sgoldstinos are assumed to be in thermal equilibrium in the early Universe, then the BBN lower bound on the gravitino mass 
becomes stronger, and of the order of 1 eV~\cite{grifols1997,gherghetta19972}.  In contrast to the BBN bounds mentioned above, 
our limit~(\ref{eq:final}) does not depend on assumptions made for the thermal bath at goldstino or sgoldstino decoupling. The cross-sections used in deriving the BBN bounds discussed here were first computed in~\cite{bhattacharya1988,bhattacharya1987}.

Given a specific model for supersymmetry mediation, our limits~(\ref{eq:final}) can be used to exclude 
a range of supersymmetry breaking scales and correspondingly also superpartner masses. 
In the simplest models, where a single parameter, e.g. the $F$-term $F_S$, determines $m_{3/2}$, 
$M_{i}$ and $m_{0}^2$, the resulting exclusion limits are not at all competitive with the corresponding 
collider limits, since they rule out supersymmetry breaking scales of the order of $10-100$~GeV, which 
have already been probed at LEP and at the LHC for a variety of models. 
Limits on the scale of supersymmetry breaking derived from~(\ref{eq:final}) remain interesting instead
for gauge mediation models with a very large number of messengers or for models which are difficult to 
probe at colliders, like those with compressed spectra, and especially for models where the gravitino mass 
$m_{3/2}$ is partially ``decoupled'' from $M_{i}$ and $m_{0}^2$, as in the examples discussed 
in Sec.~\ref{sec:theory}.

Note that the analysis at large sgoldstino couplings is unfortunately limited by the degeneracy between 
the gravitino mass and the sgoldstino coupling $\xi$, so that the prior and the range chosen for such parameter 
influence the bayesian analysis and do not allow to pinpoint the gravitino mass completely model-independently.
Therefore other constraints (or definite assumption on the SUSY breaking sector, such as those we tried
to relax) are needed to be able to draw definite statements. 

In principle our analysis could be extended to include also limits from other astrophysical objects and
that would sample a different energy range and improve the constraints. For the case of light sgoldstinos,
the inclusion of the process of sgoldstino production could bring additional constraining power,
but at the cost of a strong dependence on the unknown coupling of the sgoldstino sector.

\acknowledgments R.C. and L.C. acknowledge partial support from the European Union FP7 ITN INVISIBLES 
(Marie Curie Actions, PITN-GA-2011-289442). 
T.E. acknowledges support from the DFG research training group GRK1147.

\appendix

\section{Gravitino Luminosity}
\label{app:luminosity}
The luminosity of gravitinos produced via $\gamma(p_1)\gamma(p_2)\longrightarrow \widetilde{G}(k_1)\widetilde{G}(k_2)$ is given by~\cite{schinder1986} 
\begin{align}
L&= V \int \frac{\D^3 p_1}{(2\pi)^3\, 2p_1^0}\, 2 n_{\gamma}(p_1^0)\int \frac{\D^3 p_2}{(2\pi)^3\, 2p_2^0}\, 2 n_{\gamma}(p_2^0)  \nonumber\\
&\times \int \frac{\D^3 k_1}{(2\pi)^3\, 2 k_1^0}\int \frac{\D^3 k_2}{(2\pi)^3\, 2 k_2^0} (2\pi)^4 \delta^{(4)}(p_1+p_2-k_1-k_2)  \nonumber\\
&\times  (k_1^0+k_2^0)  \overline{\left|\mathcal{M(\gamma\gamma\rightarrow\widetilde{G}\widetilde{G})}\right|^2}\, ,
\end{align}
which corresponds to the gravitino energy emitted per unit time from a volume $V$ at a temperature $T$ via the collision of photons in thermal equilibrium. The temperature enters the luminosity via  the photon Bose-Einstein distribution function $n_{\gamma}(p_i^0)$, given by 
\begin{align*}
n_{\gamma}(p_i^0) = \frac{1}{e^{p_i^0/T}-1}\,.
\end{align*} 
In terms of the cross-section, the luminosity can be expressed as 
\begin{align}
L &=\frac{4 V}{(2\pi)^6}\int \D^3p_1  n_{\gamma}(p_1^0)\int \D^3p_2  n_{\gamma}(p_2^0) (p_1^0+p_2^0)\frac{p_1\cdot p_2}{p_1^0 p_2^0} \nonumber\\ &\times \sigma(\gamma\gamma\rightarrow\widetilde{G}\widetilde{G})\, .\label{eq:ggluminosity2}\; 
\end{align}
This expression follows from the definition of total cross-section. In order to simplify the integration, we assume $n_{\gamma}(p_i^0) > e^{-p_i^0/T}$ for all $p^0$, which is a reasonable assumption in the Supernova core~\cite{raffelt1996}. Hence,
\begin{align}
L&>\frac{4V}{(2\pi)^6}\int \D^3 p_1 \D^3p_2 e^{-(p_1^0+p_2^0)/T}(p_1^0+p_2^0)\left(1-\cos \alpha \right) \nonumber\\ &\times \sigma(\gamma\gamma\longrightarrow \widetilde{G}\widetilde{G})\, , \label{eq:ggluminosity}
\end{align}
where $\cos\alpha = \vec{p}_1\cdot \vec{p}_2/(|\vec{p}_1||\vec{p}_2|)$. In general, the cross-section depends on the photon momenta, or rather on the Mandelstam variable $s$. We express the photon momenta $p_1$ and $p_2$ in spherical coordinates $(p_i^0,\theta_i,\phi_i)$ and obtain
\begin{align}
s&= (p_1+p_2)^2 = 2 p_1^0 p_2^0 (1-\cos \alpha)\, ,\label{eq:sphoton}\\
\text{and}\; \cos \alpha &= \sin \theta_1 \sin \theta_2 \cos( \phi_1-\phi_2) + \cos \theta_1 \cos \theta_2\, . \label{eq:momentumangle}
\end{align}

\section{Invariant Amplitudes}
\label{app:amplitudes}
\subsubsection*{Gravitino Pair Production}
For the sake of completeness, we list the four invariant amplitudes relevant in the calculation of gravitino pair production via photon collision,
\begin{align}
i\mathcal{M}_{\text{Photino}} &=\frac{i}{4 M_P^2}\;\epsilon_1^{\alpha}\epsilon_2^{\beta} \; p_1^{\kappa}p_2^{\lambda}\nonumber\\
&\times \overline{\psi}^{+\,\mu}(k_2)\sigma_{\alpha\kappa}\gamma_{\mu}\frac{\slashed{q}_1-m_{\tilde{\gamma}}}{q_1^2-m_{\tilde{\gamma}}^2}\gamma_{\nu}\sigma_{\beta\lambda}\psi^{-\,\nu}(k_1) \nonumber\\
&+\frac{i}{4  M_P^2}\;\epsilon_1^{\alpha}\epsilon_2^{\beta}\;p_2^{\kappa}p_1^{\lambda} \nonumber\\
&\times \overline{\psi}^{+\,\mu}(k_2)\sigma_{\beta\kappa}\gamma_{\mu}\frac{\slashed{q_2}-m_{\tilde{\gamma}}}{q_2^2-m_{\tilde{\gamma}}^2}\gamma_{\nu}\sigma_{\alpha\lambda}\psi^{-\,\nu}(k_1) \, ,\label{eq:amp-photino}\displaybreak[0]\\
i\mathcal{M}_{\text{Graviton}} &=\frac{1}{2(p_1+p_2)^2 M_P^2}\bigg( (\epsilon_1\cdot\epsilon_2)p_1^{\lambda}p_2^{\rho} \nonumber\\
&+\frac{1}{2}\big( (p_1\cdot \epsilon_2)(p_2 \cdot\epsilon_1)-(p_1\cdot p_2)(\epsilon_1\cdot\epsilon_2)\big)\eta^{\lambda\rho}\nonumber\\
&+(p_1\cdot p_2)\epsilon_1^{\lambda}\epsilon_2^{\rho}-(p_2 \cdot \epsilon_1)\epsilon_2^{\rho}p_1^{\lambda}-(p_1\cdot \epsilon_2)p_2^{\lambda}\epsilon_1^{\rho} \nonumber\\
&+(\rho\leftrightarrow\lambda)\bigg)\nonumber\\
&\times \overline{\psi}^{+\,\mu}(k_2) \bigg[\epsilon_{\mu\sigma\nu(\lambda}\gamma^5\gamma_{\rho)}(k_2-k_1)^{\sigma} \nonumber\\
&+ \frac{i}{2}\epsilon_{\mu\sigma\nu(\lambda}\gamma^5 \left\{ \gamma^{\sigma},\sigma_{\rho)\tau}\right\}\nonumber\\
&-2im_{3/2}(2\eta_{\mu(\lambda}\eta_{\rho)\nu}-\eta_{\mu\nu}\eta_{\lambda\rho})\bigg]\psi^{-\,\nu}(k_1)\, ,\label{eq:amp-graviton}\displaybreak[0]\\
i\mathcal{M}_{\text{Scalar}} &=\frac{i c d m_{3/2}}{M_P^2} \frac{1}{2(p_1+p_2)^2-m_S^2}\nonumber\\
&\times \epsilon_1^{\alpha}\epsilon_2^{\beta} \;\left( (p_1\cdot p_2)\eta_{\alpha\beta}-p_1^{\beta}p_2^{\alpha}\right) \nonumber\\
&\times \eta_{\mu\nu}\; \overline{\psi}^{+\,\mu}(k_2)\psi^{-\,\nu}(k_1)\, ,\label{eq:amp-scalar}\displaybreak[0]\\
i\mathcal{M}_{\text{PseudoScalar}} &=-\frac{i c d}{2 M_P^2}\;\frac{1}{(p_1+p_2)^2-m_P^2} \nonumber\\
&\times \epsilon_1^{\alpha}\epsilon_2^{\beta}\,p_1^{\kappa}p_2^{\lambda}\,\epsilon_{\kappa\lambda\alpha\beta}\,(p_1+p_2)^{\zeta}\,\epsilon_{\mu\delta\nu\zeta}\;  \nonumber\\ 
&\times \overline{\psi}^{+\,\mu}(k_2)\gamma^{\delta}\psi^{-\,\nu}(k_1)\label{eq:amp-pseudoscalar}\, ,
\end{align}
where $q_i=p_i-k_2$. Here the notation is the same of~\cite{emken2013}.
\subsubsection*{Single Gravitino Production for trilinear RPV}
Here we give the invariant amplitude for single gravitino production via photon collision in presence of trilinear RPV. The effective photon-photino-neutrino vertex used in this calculation is given in~\citep{yamanaka2012}. Adopting the same notation of~\citep{yamanaka2012}, we find
\begin{align}
i\mathcal{M}_{i} &=\sum_{j}\Bigg[ \frac{\sqrt{2}}{(4\pi)^2}\frac{1}{2 M_P}\hat{\lambda}_{ijj}g_{f_j} t_n (g^{(1)}_{\tilde{f}_{j,R}}-g^{(1)}_{\tilde{f}_{j,L}}) \nonumber\\
&\times \sin \theta_{f_j}\cos \theta_{f_j} e^{i(\delta_{f_j}-\theta_n/2)}\int_0^1\text{d}z\frac{z}{zq_1^2-m^2_{\tilde{f}_{2j}}}\; \nonumber\\
&\times \epsilon^{\alpha}_1 \epsilon_2^{\beta}\left( \overline{\nu}_i(k_1)\slashed{p}_1\slashed{q}_1\gamma_{\alpha}P_L \frac{\slashed{q}_1+m_{\tilde{\gamma}}}{q_1^2-m_{\tilde{\gamma}}^2}\gamma_{\mu}\sigma_{\beta\rho}p^{\rho}_{2}\psi^{-\mu}(k_2) \right. \nonumber\\
&  + \left((p_1,\alpha)\leftrightarrow (p_2,\beta)\right)\Bigg) -(m^2_{\tilde{f}_{1j}}\leftrightarrow m^2_{\tilde{f}_{2j}} )\Bigg]\label{eq:rpvamp}
\end{align}
where $q_i=p_i-k_1$. Here $i\mathcal{M}_{i}$ is the invariant amplitude for the process $\gamma\gamma\longrightarrow \nu_i\widetilde{G}$, where the index $i$ labels the flavor of the final state neutrino. In Eq.~(\ref{eq:rpvamp}) we sum over the flavor of the fields in the loop. In addition, we denote the trilinear coupling constants by
\begin{align*}
\hat{\lambda}\equiv \begin{cases}
\lambda \quad \text{for the charged lepton-slepton loop}\, ,\\
\lambda' \quad \text{for the down-type quark-squark loop } \, .
\end{cases}
\end{align*}
and introduce the gauge coupling $g_f\equiv Q_f e$, where $Q_f$ is the electric charge of the fermion-sfermion pair in the loop. In the invariant amplitude, the additional gauge couplings $g^{(1)}_{\tilde{f}_{j,R}}$ and $g^{(1)}_{\tilde{f}_{j,L}}$ are given by
\begin{align*}
& g^{(1)}_{\tilde{q}_L} =\frac{1}{6}\frac{e}{\cos \theta_W} \, , \quad g^{(1)}_{\tilde{d}_R} = \frac{1}{3}\frac{e}{\cos \theta_W}\, , \nonumber\\ 
& g^{(1)}_{\tilde{l}_L} = -\frac{1}{2}\frac{e}{\cos \theta_W}\, , \quad g^{(1)}_{\tilde{e}_R} = \frac{e}{\cos \theta_W}\, .
\end{align*}
We also introduce the constants $t_n= Q_f n_c$, where $n_c$ is the number of colors. Finally, the sfermion mixing angles,  CP phases, and masses are donated by $\theta_{f_j}$, $\delta_{j}$, $m_{\tilde{f}_{1j}}$ and $m_{\tilde{f}_{2j}}$, respectively. 

\subsubsection*{Single Gravitino Production for bilinear RPV}
We conclude with the invariant amplitude for single gravitino production via photon collision in presence of bilinear RPV. Using the notation of \cite{emken2013}, we find
\begin{eqnarray}
i\mathcal{M} &=& \frac{g_Z \langle\tilde{\nu}\rangle U_{\tilde{\gamma} \tilde{Z}}}{8\sqrt{2}M_P m_Z} \, \epsilon_1^\alpha \epsilon_2^\beta
\bar{u}(k_2)V_{\beta}^{\gamma \nu \nu} (p_2^2) \nonumber\\
&\times& \frac{\slashed{p}_1-\slashed{k}_1+m_\nu}{(p_1-k_1)^2-m_\nu^2} (1+\gamma_5) \gamma_\nu
[\slashed{p}_1,\gamma_\alpha] \psi^{\mu}(k_1) \nonumber\\
\nonumber\\
&+& \left((p_1,\alpha) \leftrightarrow (p_2,\beta)\right) \,.
\label{eq:rpvamp2}
\end{eqnarray}
where $V_\mu^{\gamma\nu\nu}(q^2)$ is the photon-neutrino-neutrino vertex, which at one loop is given by
\begin{align}
V_\mu^{\gamma\nu\nu}(q^2) &=f_{Q}(q^2)\gamma_\mu - f_M(q^2) i\sigma_{\mu\nu}q^\nu + f_{E}(q^2)\sigma_{\mu\nu} q^\nu \gamma_5 \nonumber\\
&+ f_A(q^2) (q^2 \gamma_\mu - q_\mu \slashed{q})\gamma_5 \,.
\end{align}
The functions $f_{Q}(q^2)$, $f_{M}(q^2)$, $f_{E}(q^2)$ and $f_{A}(q^2)$ are the charge, magnetic dipole, electric dipole and anapole neutrino form factors, respectively. At zero momentum transfer, i.e. $q^2~=~0$, they are strongly constrained by observations~\cite{beringer2012}: 
\begin{eqnarray}
f_{Q}(0) &<& 10^{-13}\,, \nonumber\\ 
f_{M}(0),f_{E}(0)&<& 10^{-5}~\textrm{GeV}^{-1}\,, \nonumber\\
f_{A}(0)&<& 10^{-6}~\textrm{GeV}^{-2}\,.
\end{eqnarray}
Evaluating Eq.~(\ref{eq:vergleich1}), we assume $U_{\tilde{\gamma} \tilde{Z}}\approx m_Z/M_{1/2}$, with $M_{1/2}=\mathcal{O}(100)$~GeV.


\providecommand{\newblock}{}

\end{document}